\documentclass[a4paper,fleqn]{cas-dc}

\usepackage[authoryear,longnamesfirst]{natbib}
\usepackage[version=4]{mhchem}
\usepackage{caption}
\usepackage{midfloat}
\usepackage{dsfont}
\usepackage{hyphenat}
\usepackage{float}
\usepackage{siunitx}
\usepackage{graphicx}
\usepackage{hyperref}
\usepackage{tikz}
\usepackage{pifont}
\usepackage{placeins}
\usepackage{subcaption}
\usetikzlibrary{positioning}
\usepackage{multirow}
\usepackage[switch]{lineno}
\usepackage{balance}
\newcommand{\argmin}[1]{\underset{#1}{\text{argmin}}\,}

\def\tsc#1{\csdef{#1}{\textsc{\lowercase{#1}}\xspace}}
\tsc{WGM}
\tsc{QE}
\tsc{EP}
\tsc{PMS}
\tsc{BEC}
\tsc{DE}

\begin{document}
\let\WriteBookmarks\relax
\def\floatpagepagefraction{1}
\def\textpagefraction{.001}
\shorttitle{Data-driven ozone forecasting}
\shortauthors{D. Marvin et~al.}

\title [mode = title]{A data-driven approach to the forecasting of ground-level ozone concentration}                      

\tnotetext[1]{This work was funded by the Environment Observatory of Southern Switzerland (OASI, \hyperlink{www.ti.ch/oasi}{www.ti.ch/oasi}) of the Department of Territory of Canton Ticino (DT).}

\author[1]{Dario Marvin}
\author[1]{Lorenzo Nespoli}
\author[1]{Davide Strepparava}
\author[1]{Vasco Medici}[orcid=0000-0001-5684-4972]
\ead{vasco.medici@supsi.ch}
\cormark[1]

\address[1]{Institute for Applied Sustainability to the Built Environment, University of Applied Sciences and Arts of Southern Switzerland}
\cortext[cor1]{Corresponding author. \\\hspace*{0.6cm} Address: Via Flora Ruchat-Roncati 15, 6850 Mendrisio, Switzerland}

\begin{abstract}
The ability to forecast the concentration of air pollutants in an urban region is crucial for decision-makers wishing to reduce the impact of pollution on public health through active measures (e.g. temporary traffic closures). In this study, we present a machine learning approach applied to the forecast of the day-ahead maximum value of the ozone concentration for several geographical locations in southern Switzerland. 
Due to the low density of measurement stations and to the complex orography of the use case terrain, we adopted feature selection methods instead of explicitly restricting relevant features to a neighbourhood of the prediction sites, as common in spatio-temporal forecasting methods. We then used Shapley values to assess the explainability of the learned models in terms of feature importance and feature interactions in relation to ozone predictions; our analysis suggests that the trained models effectively learned explanatory cross-dependencies among atmospheric variables. Finally, we show how weighting observations helps in increasing the accuracy of the forecasts for specific ranges of ozone's daily peak values.

\end{abstract}


\begin{keywords}
Shapley values \sep Genetic algorithms \sep Environmental forecasting \sep Evaluating forecasts\sep Multivariate time series \sep  
\end{keywords}
\maketitle
\section{Introduction and motivations}
Ground-level ozone (\ce{O3}), which is formed in the troposphere by photochemical reactions in the presence of sunlight and precursor pollutants, such as the oxides of nitrogen (\ce{NO_x}) and volatile organic compounds (VOCs) \citep{calvert2015mechanisms}, is known to be a pollutant particularly dangerous to human health \citep{WHO2003,Stewart2017}. The forecast of ozone concentrations is an important task to ensure the protection of outdoor workers who are exposed to polluted air during the most dangerous hours of the day, as well as sensitive people such as children or the elderly.\\
In this paper, we tested a number of machine learning algorithms that forecast the maximum hourly ozone concentration of a given day by performing the prediction at two different times: the evening before and the early morning of the target day. We are particularly interested in the days in which the ozone concentration is significantly higher than usual, due to their potential impact on public health. The choice of the target variable is based on Swiss legislation, which states that the 1-hour mean of $120 \mu g/m^3$ must not be exceeded more than once per year \citep{OAPC}.
We use an empirical, data-centric approach that leverages a large data set of air quality, weather station measurements and weather forecasts. Data are collected for seven sites in southern Switzerland, for which the forecast is performed.\\
We use Shapley values and a genetic algorithm to select the most important features for each model, and we calculate various forecasts with the help of cutting-edge forecasting algorithms.\\
The work is organized as follows: Section \ref{sec:related_works} contains an overview of similar research in the scientific literature, while Section \ref{sec:contributions} highlights the contributions we bring to this work. Section \ref{sec:data} introduces the used dataset and the nomenclature we used trough the paper to refer to the different features, while Section \ref{sec:formulation} presents the forecasting problem peculiarities and the problem formulation. Section \ref{sec:FS} describes the two feature selection methods that have been tested, namely a custom genetic algorithm and a feature selection based on Shapley values. Section \ref{sec:algo} outlines the regression algorithms used to perform the analysis. Section \ref{sec:kpi} introduces the deterministic and probabilistic KPIs that we have used to evaluate the different forecasters. Section \ref{sec:NumFeat} presents the results of the two tested feature selection algorithms. In Section \ref{sec:feat_imp} we study how different features and features' interactions affect the final predictions of the forecast, using Shapley values. In Section \ref{sec:results}, we show the numerical results for the tested forecasting algorithms, while in Section \ref{subsec:high_peaks} we focus on the prediction of extreme events.
Finally, Section \ref{sec:conclusions} concludes the paper with a summary of our main findings.

\subsection{Related works} \label{sec:related_works}
Tropospheric ozone concentration has been the subject of several studies, both for prediction (the task of finding a map from a set of covariates to a target) and for forecasting (predicting the values of the target in advance, in future time steps). 
In \cite{AlAbri2015} different non-parametric models from the WEKA toolkit are tested to derive the ozone concentration from a set of 8 different gaseous chemicals and atmospheric conditions measured at a single location. Similarly, WEKA is used in \cite{Mohan2019} to adapt models representing atmospheric conditions to the ozone concentration at ground level, which shows that even summer ozone peaks can be accurately predicted if the atmospheric conditions are known. In \cite{Feng2011}, meteorological data from a site near Beijing were used to predict the hourly ozone concentration at that point by using a neural network whose weights were trained using a genetic algorithm. In addition, different models were adapted for different times of the day. In \cite{Sheta2018}, a nonlinear state-space model using \ce{PM10}, temperature, wind speed, and relative humidity as input is identified by using a neural network to predict ozone concentration. The model is then compared with linear models and multilayer perceptron. In \cite{Siwek2016} the authors used a data set of 55 characteristics (meteorological conditions and their statistical transformations) collected in Warsaw to predict various air pollutants. They showed how by reducing the number of features with a pre-selection step, the final accuracy of the prediction could be increased. Two pre-selection methods were compared, one with a genetic algorithm and a stepwise greedy strategy for linear models.

The task of forecasting PM$_{2.5}$ and ozone concentrations for three large Chinese cities is considered in  \cite{Lv2016}. Like in our study, the authors considered multiple monitoring stations, but the final values of the relevant atmospheric variables were weighted averages of neighbours of the target cities. The forecasts have been obtained by fitting knowledge-based empirical formulae using historical data. No systematic investigation of variables interaction has been carried out. The authors showed how the maximum daily temperature is the single most relevant variable in predicting (and forecasting) ozone concentration. They consistently found a strong correlation in numerical weather prediction forecasts error of this value with the error for ozone prediction. In \cite{Eslami2019} the authors proposed a deep convolutional neural network (CNN) to forecast the hourly ozone concentrations for the day-ahead, over 25 monitoring sites. Despite the ability of the CNN in correctly predicting daily ozone trends, the authors found that it under-predicted high ozone peaks during the summer. In \cite{Gong2016}, the authors focused on the forecast of extreme ozone concentrations, which are also the most useful to predict. Forecasting extreme events is, in fact, more complicated than predicting them, as found out in \cite{Mohan2019}. When one is mostly interested in predicting these tail events, sampling techniques can be applied in order to mitigate the class imbalance problem (rare events are under-represented in the training data). The authors of this study applied different sampling methods to increase classification accuracy of ozone concentration, considering three different classes. They found out that under-sampling can indeed increase the classification performance. Unfortunately, a drawback of this technique is that several data of the most represented class are discarded, which could lead to a lack of generalization of the model, due to over-fitting or a reduction of cross-learning (learning patterns from data in a given class, which are also present in a second unobserved class, which could increase the prediction accuracy).   

\subsection{Contributions} \label{sec:contributions}
In the presence of a high number of relevant features, the task of forecasting the next day peak in the concentration of ozone becomes highly challenging, due to the low number of observations on which a forecasting algorithm can be trained. In fact, having a dataset consisting of a few years of observations could result in having a number of features higher than the number of observations, as in the presented case. On the other hand, observations further back in time may not be representative of the current situation, as the mixture of nitrogen oxides in the air has changed over time following vehicle fleet renewal. As a consequence, due to the scarce number of instances, we couldn't apply under-sampling techniques, as done in \cite{Mohan2019}. The only effective way of training a model is by applying some dimensionality reduction techniques. Our first contribution consists of evaluating two different methods to perform feature selection. First, we tested a genetic algorithm, as was done in \cite{Siwek2016} for the pollutant prediction task. In this case, we crafted custom mutation and crossover functions tailored to the forecasting task. The second approach we tested is based on Shapley values \citep{Lundberg2017}. We then evaluated and compared the two feature selection methods. To show that the feature selection step is beneficial in increasing the accuracy of the predictive algorithm, we compared our models with two control cases: one in which the model uses all the available features and one in which we pick the features entirely at random. Our second contribution is to compare the performances of different popular learning algorithms trained on the selected features. Thirdly, we investigate the effect of imposing weights on the observations with the highest daily ozone concentration on the algorithms' forecasting quality of extreme values. Our final contribution is an a-posteriori explanation of feature importance. We investigate the more relevant feature interactions in predicting the ozone peak and explain our findings in terms of atmospherics's physics.

\section{Dataset} \label{sec:data}
\subsection{Geographical context and data acquisition}
\begin{figure*}
    \centering
    \includegraphics[width=\textwidth]{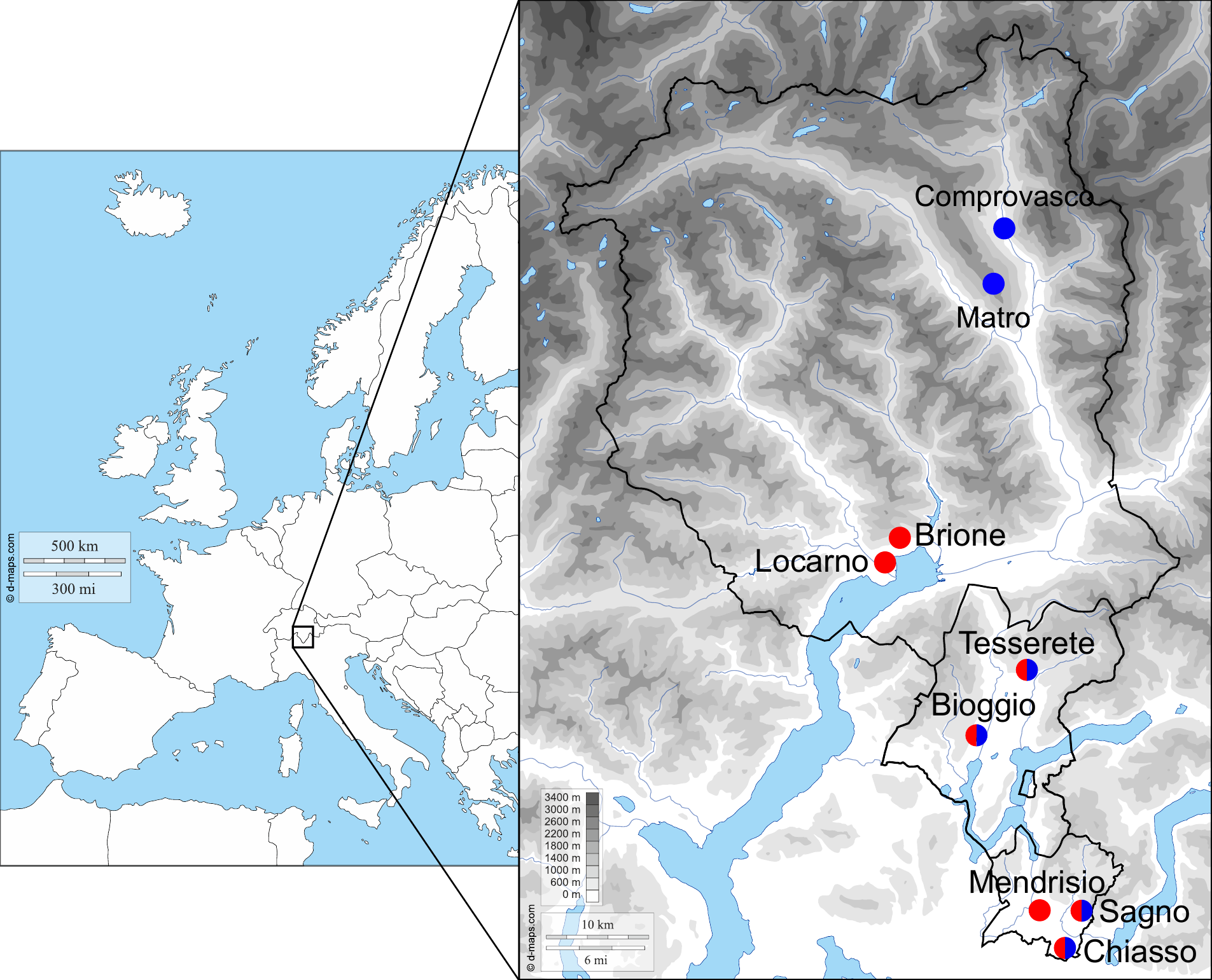}
    \caption{Map of Ticino with the location of the stations used. Red shows the stations where air quality and weather measurements are collected and blue shows the locations for which weather forecasts are available. The maps were originally downloaded from \cite{MapEurope} and \cite{MapTicino}.}
    \label{fig:ticino}
\end{figure*}

In this study, we focused on the Canton of Ticino, the southernmost canton of Switzerland. In this region, the concentration of air pollutants is generally higher than in the rest of the country and is influenced both by the orography and the level of urbanization and industrialization. The natural shield provided by the Alps makes Ticino the region with the highest solar radiation rate in Switzerland. Ticino is characterized by a densely populated and heavily trafficked southern region, and by a sparsely populated and more mountainous northern region. It also borders Lombardy to the south, the most industrialized region in Italy.\\
In this study, we used data acquired from several air quality and weather stations distributed in the region. In addition, the Swiss Federal Office of Meteorology and Climatology MeteoSwiss\footnote{https://www.meteoswiss.admin.ch/} numerical weather prediction (NWP) service provided weather forecasts for some of these locations. Fig. \ref{fig:ticino} shows the position of the monitoring stations and the locations for which the weather forecasts are available. Tables \ref{tab:stations} and \ref{tab:forecast_station} describe the geographical context for monitoring stations and the weather forecasting locations, respectively.

\begin{table}[pos=htbp]
\centering
\caption{Geographic context of air quality and weather monitoring stations}
\setlength{\tabcolsep}{0.5em} 
{\renewcommand{\arraystretch}{1.2}
\begin{tabular}{|l|c|c|l|l|}
\hline
\multicolumn{1}{|c|}{\textbf{Station}} & \multicolumn{1}{c|}{\textbf{Code}} & \multicolumn{1}{c|}{\textbf{\vtop{\hbox{\strut Altitude}\hbox{\strut [m a.s.l.]}}}} & \multicolumn{1}{c|}{\textbf{Context}}  & \multicolumn{1}{c|}{\textbf{\vtop{\hbox{\strut Main \ce{O3}}\hbox{\strut \, source}}}} \\ \hline
Locarno   & $l_1$  & 200  & Urban    & Industry      \\
Brione    & $l_2$  & 486  & Suburb & Valley floor  \\
Bioggio   & $l_3$  & 314  & Suburb & Industry      \\
Tesserete & $l_4$  & 518  & Rural    & Valley floor  \\
Chiasso   & $l_5$  & 230  & Urban    & Industry      \\
Mendrisio & $l_6$  & 354  & Suburb & Industry      \\
Sagno     & $l_7$  & 704  & Rural    & Valley floor  \\ \hline
\end{tabular}}
\label{tab:stations}
\end{table}

\begin{table}[pos=htbp]
\centering
\caption{Geographic context of weather forecasting locations}
\label{tab:forecast_station}
\setlength{\tabcolsep}{0.5em} 
{\renewcommand{\arraystretch}{1.2}
\begin{tabular}{|l|c|c|l|l|}
\hline
\multicolumn{1}{|c|}{\textbf{Location}} & \multicolumn{1}{c|}{\textbf{Code}} & \multicolumn{1}{c|}{\textbf{\vtop{\hbox{\strut Altitude}\hbox{\strut [m a.s.l.]}}}} & \multicolumn{1}{c|}{\textbf{Context}}        \\ \hline
Comprovasco   & $p_1$ & 575 & Rural \\
Matro         & $p_2$ & 2171 & Mountain\\
Bioggio       & $p_3$ & 518 & Suburb   \\
Tesserete     & $p_4$ & 626 & Rural    \\
Chiasso       & $p_5$ & 240 & Urban    \\
Sagno         & $p_6$ & 704 & Suburb   \\ \hline
\end{tabular}}
\end{table}

Due to its photochemical origin, \ce{O3} shows a strong seasonal pattern, with higher concentrations in summer. For this reason we have focused our analysis only on the period from May to September in the years between 2015 and 2019. This period was chosen to take into account a significant number of measurements, i.e. enough to train the algorithms correctly, while simultaneously avoiding the use of previous years, when emissions of precursors \ce{NO}, \ce{NO2} and \ce{NO_x} in Switzerland were more intense than today. We used the data from the first 4 years to train the forecasting algorithm and the data from 2019 to test them.\\
A variety of signals covering weather and air quality with hourly resolution were considered for the model, as shown in Table \ref{tab:signals}. Most monitoring stations record the full set of signals specified in Table \ref{tab:signals} on site. The few exceptions, where some signals were not collected locally, were managed with data from the nearest available stations. When training forecasting algorithms, we considered data measured up to 72 hours in the past (except for the ARIMAX model) and weather forecasts up to 33 hours in the future. The 72 hours value was chosen based on preliminary results, in which we considered a history length up to 7 days, and systematically shorten it. As we didn't experience significant accuracy improvement using a history length higher than 72 hours, we have fixed this value for all the experiments. Table \ref{tab:signals} shows the available signals.

\begin{table}[pos=htbp]
\centering
\setlength{\tabcolsep}{0.5em} 
\renewcommand*{\arraystretch}{1.2}
\caption{Dataset description. The symbol $\dagger$ denotes a variable that is both measured and forecasted by the NWP service, while the symbol $\ddagger$ indicated a forecasted only signal.}
\label{tab:signals}
\begin{tabular}{|l|l|l|}
\hline
\multicolumn{1}{|c|}{\textbf{Signal}}  & \multicolumn{1}{c|}{\textbf{Symbol}} & \multicolumn{1}{c|}{\textbf{Unit}} \\ \hline
Nitrogen oxide & $NO$ & $\left[\SI[per-mode=repeated-symbol]{}{\micro\gram\per\cubic\meter}\right]$ \\
Nitrogen dioxide & $NO2$ & $\left[\SI[per-mode=repeated-symbol]{}{\micro\gram\per\cubic\meter}\right]$ \\
Generic nitrogen oxides & $NOx$ & $\left[\SI[per-mode=repeated-symbol]{}{ppb}\right]$ \\
Ozone & $O3$ & $\left[\SI[per-mode=repeated-symbol]{}{\micro\gram\per\cubic\meter}\right]$ \\
$\dagger$ Global irradiance & $G$ & $\left[\SI[per-mode=repeated-symbol]{}{\watt\per\square\meter}\right]$ \\ 
$\dagger$ Atmospheric pressure & $P$ & $\left[\SI[per-mode=repeated-symbol]{}{\hecto\pascal}\right]$ \\
$\dagger$ Precipitation & $Prec$ & $\left[\SI[per-mode=repeated-symbol]{}{\milli\meter}\right]$ \\
$\dagger$ Relative humidity & $RH$ & $\left[\SI[per-mode=repeated-symbol]{}{\percent}\right]$ \\
$\dagger$ Temperature & $T$ & $\left[\SI[per-mode=repeated-symbol]{}{\celsius}\right]$ \\
$\ddagger$ Dew point & $TD$ & $\left[\SI[per-mode=repeated-symbol]{}{\celsius}\right]$ \\
$\dagger$ Wind direction, vectorial average & $Wd$ & $\left[\SI[per-mode=repeated-symbol]{}{\degree}\right]$ \\
$\dagger$ Wind speed, vectorial average & $Ws$ & $\left[\SI[per-mode=repeated-symbol]{}{\meter\per\second}\right]$ \\
$\dagger$ Cloud cover & $CN$ & [-] \\ \hline
\end{tabular}
\end{table}

For various reasons, such as station maintenance, data transmission failure or power outages, part of the data in the time series of the years from 2015 to 2019 are missing. During the training period, data completeness is above 99\%, but two stations have substantial data holes in 2019: in Tesserete and Sagno respectively, 50 and 15 full days of measurements are missing during the test period. All the missing data in the training set have been filled using a random forest with surrogate splits, trained to predict the missing data using the station itself and its neighbours.

\subsection{Feature engineering}
The large number of signals and their high granularity resulted in a high dimensionality of the dataset. To reduce the overall number of features and minimize the computational effort, we partly replaced the hourly values of the measured and forecasted signals with basic statistical aggregations, i.e. minimum, maximum and average value over a longer time period, as illustrated in Table \ref{tab:chunks}.

\begin{table*}[pos=htbp]
\centering
\caption{Summary of all the features used in this study. More information about the \textit{MOR} and \textit{EVE} cases is given in Section \ref{sec:nomenclature}}
\label{tab:chunks}
\setlength{\tabcolsep}{0.5em} 
{\renewcommand{\arraystretch}{1.2}
\begin{tabular}{|l|l|c|l|} 
\hline
\multicolumn{1}{|c|}{\textbf{Signal kind}}                                                                                   & \multicolumn{1}{c|}{\textbf{Time interval}}                                                                                                                                                                                                               & \multicolumn{1}{c|}{\textbf{Code}}                                                                   & \multicolumn{1}{c|}{\textbf{Aggregation}}                                                                                                  \\ 
\hline
\multirow{4}{*}{All Measured data} & Past 24 hours ($m_0, \ldots, m_{23}$) & $m_i$                     & Hourly values                    \\ 
\cline{2-4} 
                                   & From 0 to 24 hour before            & $24h$                     & \multirow{3}{*}{Mean}            \\
                                   & From 0 to 48 hour before            & $48h$                     &                                  \\
                                   & From 0 to 72 hour before            & $72h$                     &                                  \\ \hline
\begin{tabular}[c]{@{}l@{}}All forecasts\\ (same station) \end{tabular}                       & \begin{tabular}[c]{@{}l@{}}MOR: from $s_0$ to $s_{32}$ \\ EVE: from $s_0,$ to $s_{29}$ \end{tabular}                                                                                                                                                                                         & $s_i$                                                                  & Hourly values                                                                                                \\ 
\hline
\multirow{4}{*}{\begin{tabular}[c]{@{}l@{}}All forecasts\\ (neighbour station) \end{tabular}} & \begin{tabular}[c]{@{}l@{}}MOR: from $s_0$ to $s_{7}$ \\ EVE: from $s_0,$ to $s_{6}$ \end{tabular}                                                                                                                         & $b_1$                                                                  & \multirow{4}{*}{\begin{tabular}[c]{@{}l@{}}Minimum, maximum and\\ average of every bin $b_i$ \end{tabular}}   \\ 
\cline{2-3}
                                                                                              & \begin{tabular}[c]{@{}l@{}}MOR: from $s_8$ to $s_{16}$\\ EVE: from $s_{7}$ to $s_{13}$ \end{tabular}                                                                                                                       & $b_2$                                                                  &                                                                                                              \\ 
\cline{2-3}
                                                                                              & \begin{tabular}[c]{@{}l@{}}MOR: from $s_{17}$ to $s_{24}$\\ EVE: from $s_{14}$ to $s_{19}$ \end{tabular}                                                                                                                    & $b_3$                                                                  &                                                                                                              \\ 
\cline{2-3}
                                                                                              & \begin{tabular}[c]{@{}l@{}}MOR: from $s_{25}$ to $s_{32}$\\ EVE: from $s_{20}$ to $s_{29}$ \end{tabular}                                                                                                                    & $b_4$                                                                  &                                                                                                              \\ 
\hline
measured $NOx$                                                                                & \begin{tabular}[c]{@{}l@{}}MOR: previous afternoon ($m_6$ to $m_{18}$)\\EVE: previous morning ($m_6$ to $m_{18}$) \end{tabular}                                                                                             & \multicolumn{1}{c|}{$NOx_{12h}$}                                       & Mean                                                                                                         \\ 
\hline
Forecasted $T$                                                                                & \begin{tabular}[c]{@{}l@{}}MOR: upcoming afternoon ($s_6$ to $s_{18}$)\\EVE: upcoming afternoon ($s_{18}$ to $s_{29}$) \end{tabular}                                                                                        & \multicolumn{1}{c|}{$\widehat{T}_{PM}$, $\widehat{T}_{PM,squared}$ } & Mean and squared mean                                                                                        \\ 
\hline
Forecasted $T$                                                                                & \begin{tabular}[c]{@{}l@{}}MOR: all hourly values, from $s_0$ to $s_{32}$\\EVE: all hourly values, from $s_0$ to $s_{29}$ \end{tabular}                                                                                                                                                               & \multicolumn{1}{c|}{$\widehat{T}_{max}$}                               & Maximum                                                                                                          \\ 
\hline
Forecasted $TD$                                                                                & \begin{tabular}[c]{@{}l@{}}MOR: all hourly values, from $s_0$ to $s_{32}$\\EVE: all hourly values, from $s_0$ to $s_{29}$ \end{tabular}                                                                                                                                                                                    & \multicolumn{1}{c|}{$\widehat{TD}_{max}$, $\widehat{TD}_{max}^{transf}$}                               & Maximum, $(\text{Maximum} +20)^3$                                                                                                          \\ 
\hline
Forecasted $G$                                                                                & \begin{tabular}[c]{@{}l@{}}MOR: upcoming morning ($s_0$ to $s_{6}$)\\MOR: upcoming afternoon ($s_6$ to $s_{18}$)\\EVE: upcoming morning ($s_{6}$ to $s_{18}$)\\EVE: upcoming afternoon ($s_{18}$ to $s_{29}$)\end{tabular} & \multicolumn{1}{c|}{$\widehat{G}_{AM}$, $\widehat{G}_{PM}$}          & Mean                                                                                                         \\ 
\hline
Forecasted $CN$                                                                               & \begin{tabular}[c]{@{}l@{}}MOR: upcoming morning ($s_0$ to $s_{6}$)\\MOR: upcoming afternoon ($s_6$ to $s_{18}$)\\EVE: upcoming morning ($s_{6}$ to $s_{18}$)\\EVE: upcoming afternoon ($s_{18}$ to $s_{29}$)\end{tabular} & \multicolumn{1}{c|}{$\widehat{CN}_{AM}$, $\widehat{CN}_{PM}$}        & Mean                                                                                                         \\ 
\hline
Forecasted $Prec$                                                                             & Upcoming 24 hours ($s_0$ to $s_{23}$)                                                                                                                                                                                       & \multicolumn{1}{c|}{$\widehat{Prec}_{24h,sum}$}                        & Sum                                                                                                          \\
\hline

measured $YO3$                                                                             & \ce{O_3} measurements of the previous day                                                                                                                                                                                     & \multicolumn{1}{c|}{$YO3$}                        & Maximum                                                                                                          \\
\hline
forecasted $RHW$                                                                             & One categorical value for the prediction day                                                                                                                                                                                       & \multicolumn{1}{c|}{$RHW$}                        & -                                                                                                          \\
\hline
\end{tabular}}
\end{table*}

Based on suggestions from experts in the field of atmospheric physics\footnote{Environment Observatory of Southern Switzerland (OASI)}, we further manipulated some of the signals available in the dataset to create additional features. The engineered features are listed in Table \ref{tab:chunks}. In addition, we also included a categorical feature, called $RHW$, which describes the general situation of the weather in Switzerland for the prediction day using 12 weather types. 

For each location, separate forecasting models are trained using a subset of the matrix of all features. This subset contains the data specific to the location and information from the neighboring stations. For NWP, hourly values are used for the specific location, while bins are used for the data of the neighboring stations, as summarized in Table \ref{tab:chunks}. For example, the dataset of Chiasso contains the hourly NWP from Chiasso itself and bins aggregations from Sagno.\\
Given the different number of stations involved each time, the number of features for each model is variable and comprised between 1700 and 2100.

\subsection{Nomenclature} \label{sec:nomenclature}
\begin{figure*}
\centering  
\includegraphics[width=0.95\textwidth]{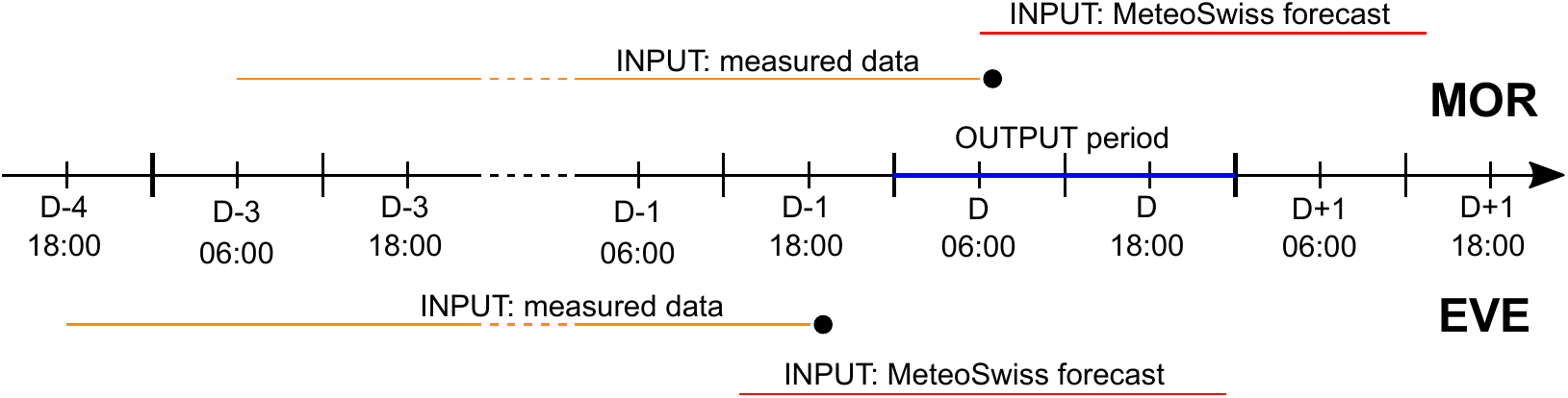}
\caption{Time window of input and output data. Times are intended in local time (CEST).}
\label{fig:time_window}
\end{figure*}

The ozone forecast at any station for any given day $\text{D}$ is computed twice: the first time at 18:30 (16:30 UTC) of the previous day $\text{D}-1$, which we call \textit{EVE} forecast, and the second time at 06:30 (04:30 UTC) of the same day $\text{D}$, here called \textit{MOR} forecast. This is because the weather forecasts issued by the NWP services are published twice a day, at 05:00 and at 14:00 local time. Fig. \ref{fig:time_window} illustrates the time window for a generic day. For each station, we tested 8 different prediction methods at the respective prediction times \textit{EVE} and \textit{MOR}, for a total of 16 models per station.\\

When labeling the aggregated data in Table \ref{tab:chunks}, we use the following conventions. Measured quantities are denoted by the letter $m$ and weather forecasts provided by NWP services are denoted by the letter $s$. The index is the difference in hours between the last available data point and the acquisition time. Following this convention, $m_0$ refers to the last measured data point available, i.e., the value measured at 06:00 for \textit{MOR} and at 18:00 for \textit{EVE}. $m_1$ refers to the value measured at 05:00 and 17:00 respectively, and so on up to $m_{23}$.\\
The same temporal indexing applies to values provided by NWP services. For \textit{MOR} we call $s_0$ the forecasted value at 06:00, $s_1$ the value for 07:00 and so on. In the \textit{EVE} case we call $s_0$ the predicted value at 18:00, $s_1$ the predicted value at 19:00 and so on. The structure of the aggregation bins is shown in Table \ref{tab:chunks}.\\
To better refer to each specific component of the models, we denote the features based on the location of the measurement and time to which it is referred, combining the codes of Tables \ref{tab:signals}, \ref{tab:chunks},  \ref{tab:stations} and \ref{tab:forecast_station}. For example, $G_{m_1}^{l_1}$ designates the global irradiance measured in Locarno at 05:00 in the \textit{MOR} model and at 17:00 in the \textit{EVE} model. Similarly, $\widehat{T}_{s_{10}}^{p_3}$ is the forecasted temperature in Bioggio at 16:00 in the \textit{MOR} model, and at 04:00 of the following day in the \textit{EVE} model. $NO_{72h}^{l_2}$ is the mean value of all the measured NO concentrations up to 72 hours before the prediction, in Brione.

\section{Problem formulation} \label{sec:formulation}

The problem of forecasting the daily maximum ozone signal presents the following characteristics:
\begin{enumerate}
    \item The signal is strongly seasonal due to the presence of annual patterns in both anthropogenic and non-anthropogenic processes governing the ozone generation.
    \item The signal is non-stationary since its variance is subject to inter- and intra-annual fluctuations.
    \item The forecasts dependence on the features is non-linear, as described in the literature and as further detailed in Section \ref{sec:feat_imp}.
    \item The forecasted values are physically bounded by the photo-chemistry and advective phenomena regulating the formation and transport of ozone in the troposphere and atmosphere.
    \item In our use case, monitoring stations providing measurements of relevant features for the ozone forecasting, such as temperature, past ozone and NOx values, are not dense enough (nor at similar distances from prediction points) to provide a regular mesh, as can be seen in Fig. \ref{fig:ticino}. In this case, the use of spatio-temporal Gaussian processes \cite{Kupilik2018}, Gaussian Markov Random Fields \cite{Cameletti2013} or other graph-based spatio-temporal techniques \cite{Carrillo2020} can lead to poor results.
\end{enumerate}  
Given the above considerations, we chose to model daily ozone maxima using separate predictors for each location, while still taking into account relevant features from nearby locations. The neighbouring stations for each prediction point are illustrated in Fig. \ref{fig:ticino}, where the whole region is divided into three macro zones. In any case, we let the feature selection processes described in Section \ref{sec:FS} discriminate whether a given measurement station is relevant or not.  Calling $n$ the number of observations and $k$ the number of features, we define a training and a test dataset, $\mathcal{D}_{tr} = \{  x_{tr}, y_{tr}\}$ and $\mathcal{D}_{te} = \{  x_{te}, y_{te}\}$, respectively, where $x_{tr} \in \mathds{R}^{n_{tr} \times k}$, $x_{te} \in \mathds{R}^{n_{te} \times k}$,  are matrices of features, and $y_{tr} \in \mathds{R}^{n_{tr}}$ and $y_{te} \in \mathds{R}^{n_{te}}$ are the targets' vectors, containing the maximum hourly ozone concentration of the same days. In this paper, $n_{tr}$ and $n_{te}$ are equal to 587 and 151, respectively, that is, the number of available days between May and September for the 2015-2018 period and for 2019. On the other hand, the number of features, $k$, has been kept fixed to $30$ for all the numerical experiments and raised to 100 for the study on the high peaks prediction in Section \ref{subsec:high_peaks}. These numbers were chosen experimentally by systematically increasing them and choosing the $k$ value beyond which the predictors' performance no longer increased significantly.

We train a model $f(x_{tr},\Theta)$, where $\Theta$ is a set of the model's parameters, in order to produce the forecasts for unseen data $x_{te}\in\mathds{R}^{n_{te}\times k}$:
\begin{linenomath*}
\begin{equation}
    \hat{y}_{te} = f(x_{te},\Theta).
\end{equation}
\end{linenomath*}
In order to compare the results across the different approaches, we used regression specific key performance indicators (KPIs), as classification scores can only be compared while using the same bins for the choice of the classes. Different values for the bins' edges are used in the ozone prediction literature since those are typically chosen based on the local legislation. As such, we trained the model $f(x_{tr},\Theta)$ minimizing L2 loss:
\begin{linenomath*}
\begin{equation}
    \Theta^* = \argmin{\Theta} \left\Vert y_{tr} - f(x_{tr},\Theta)\right\Vert_2^2.
\end{equation}
\end{linenomath*}
We highlight how this notation must be slightly adapted for the ARIMAX model introduced in Section \ref{sec:algo}; in this case the model can be described as $f(x_{tr},y_{tr},\Theta)$, where the endogenous input signal $y_{tr}$ is then opportunely shifted with the use of the backshift operator, as further explained in Section \ref{sec:algo}.

\section{Feature selection methods} \label{sec:FS}
Given the large number of features in each model, if we were to train the prediction algorithms using all the variables, whose number largely exceeds the number of available observations, we could potentially incur numerical problems of solution non-uniqueness and multicollinearity that would corrupt the prediction process. Moreover, even if the dataset contained a proportional number of observations, an excessive number of features would still result in a long computational time, which is justified only if the forecasting performance is better than that of an algorithm trained on a subset of the features.\\
In this paper, we decided to perform a feature selection using a custom implementation of a genetic algorithm (GA), as well as using a proceeding issued from game theory, exploiting Shapley values. The effectiveness of these two approaches is compared in Section \ref{sec:NumFeat} against a model composed of features picked at random and a model composed of all the available features.\\

\subsection{Feature selection using genetic algorithm} \label{sec:GA}
In our implementation of the GA, an individual $A$ is defined as a subset of the entire feature set $F$ with cardinality k
\begin{linenomath*}
\begin{equation}
A \subset F, \quad \left|A\right|=k
\end{equation}
\end{linenomath*}
where k is the number of retained features. As anticipated in Section \ref{sec:formulation}, in this study we set $k=30$.\\
We have defined a crossover function that ensures that the offsprings that emerge from it still retain k features from their parents, with no repetitions. Formally, the offspring $C$ is a subset of the union of the sets of its parents $A$ and $B$, with cardinality $k$:
\begin{linenomath*}
\begin{equation}
C \subset \left(A \cup B \right), \quad \left| C\right| = k
\end{equation}
\end{linenomath*}
We defined a custom mutation function so that each feature of the offspring $C$ is either the original feature of its parent $A$ with probability 95\%, or a new feature from $B \subset (F \setminus A)$ with probability 5\%, and such that $C$ has only unique features. In practice, we generate two sequences $a$ and $b$ from the sets $A$ and $B$, by randomly fixing their order and iterate on them to generate the new set $C$, which is composed by the elements of the sequence c, where the $i^{\text{th}}$ element of c is defined as:
\begin{linenomath*}
\begin{equation}
c_{i} = a_{i}(u \geq 0.05)+b_{i}(u<0.05) \quad  u \sim U[0,1]\\
\end{equation}
\end{linenomath*}
The fitness function is defined as the out-of-bag mean squared error (MSE) of a random forest composed of 30 bootstrap-aggregated (bagged) decision trees, trained on the $k$ active features of the individuals.
We selected a population size of $10k$ individuals, a crossover fraction of 80\%, and an elite count of 5\% of the population size. The GA stops after 100 stall generations and the feature set of the best individual is selected.

\subsection{Feature selection using Shapley values} \label{sec:SHAP}
Another method to assess the importance of each feature is presented in \cite{Lundberg2017}. This method assigns feature importance scores using Shapley values, which originated in the field of game theory where they are used to estimate the contribution of various agents in increasing the welfare of a community. These are expressed as:
\begin{linenomath*}
\begin{equation}
\phi_{i}(f, x_{tr})=\sum_{z \subseteq x} \frac{\left|z\right| !\left(k-\left|z\right|-1\right) !}{k !}\left[f\left(z,\Theta \right)-f\left(z \backslash i,\Theta\right)\right]
\end{equation}
\end{linenomath*}
where $f(x_{tr},\Theta)$ is a regression model, in our specific case the NGBoost algorithm, which will be introduced in Section \ref{sec:algo}, $x_{tr}$ is the feature set on which the model has been trained, $\Theta$ is a model-specific set of parameters, $k$ is the number of variables in the training set $x_{tr}$, and $z \backslash i$ denotes the minus set operation, that is, the subtraction of the $i_{th}$ feature from the reduced dataset $z$. The authors in \cite{Lundberg2017} have shown that such coefficients have highly desirable properties, which favourably affect their ability in the (local) explanation of the models, and have been shown to be consistent and more robust with respect to other more widespread methods for the evaluation of feature importance. Furthermore, authors in \cite{Lundberg2020} recently proposed a computationally efficient algorithm specifically tailored for tree-based models, and made it available through the \texttt{shap} python package. The \texttt{shap} package provides an exact computation of Shapley value explanations for tree-based models. This provides local explanations with theoretical guarantees of local accuracy and consistency, which increase the robustness of the method, since it doesn't rely on random samplings, which would be required to find the Shapley values using approximate algorithms.

\section{Regression models} \label{sec:algo}
After the $k$ features that best explain the data have been selected, we use them to create the regression matrix and produce the test forecasts. For this work, we studied the output of several parametric models, such as linear regression, Ridge regression, LASSO and ARIMAX, as well as more complex non-parametric tree-based algorithms, such as random forests, XGBoost, NGBoost and LSBoost, described below. 

\paragraph{\textbf{Penalized linear regression algorithms}}
Ridge regression is a method designed to avoid collinearity issues and avoid near-singular matrix inversions when solving linear regression problems, especially in the case in which the number of features is large compared to the number of observations. In this case, the regression coefficients $\beta \in \mathds{R}^k$ are quadratically penalized with parameter $\lambda$, such that the closed form solution becomes: 
\begin{linenomath*}
\begin{equation}
    \hat{\beta}_R = \left(x_{tr}^Tx_{tr}+\lambda I_k\right)^{-1}x_{tr}^{T}y_{tr},
\end{equation}
\end{linenomath*}
where $I_k$ is the identity matrix of size $k$. In this work, $\lambda$ is tuned in cross validation on the training set. 
Instead of punishing $\beta$ using the L2 norm, the \textit{LASSO} (Least Absolute Shrinkage and Selection Operator) regression \citep{Tibshirani1996} penalizes $\beta$ using the L1 norm, such that some of the elements of $\hat{\beta}$ could be set to zero. Unlike Ridge regression, LASSO does not have a closed form solution, but it has to be approximated through numerical methods.

\paragraph{\textbf{ARIMAX}}
The well known Autoregressive Integrated Moving Average with explanatory variable (ARIMAX) is defined as:
\begin{linenomath*}
\begin{equation}
    \hat{y}_t = \beta x_{tr} + \sum_{i=1}^p \phi_i B^i(y'_t) - \sum_{i=1}^q \theta_i \epsilon_{t-1} + \epsilon_t.
\end{equation}
\end{linenomath*}
where $B$ is the backshift operator, i.e. $B^n(z_t) = z_{t-n}$, $\epsilon_t$ is an additive white Gaussian noise and the time series $y'$ is the result of differencing $y_{tr}$ $d$ times. The matrix $x_{tr}$ contains the daily values of the selected features up to the day before the prediction and $\beta$ are the regression coefficients as usual.\\
The models are created and calculated with \texttt{statsmodels}'s SARIMAX function and the parameters $p, q, d$ are chosen via grid search and fitted using maximum likelihood estimation; we considered only $d = 0$ since we don't have important trends, while we set 7 and 3 as maximum values for $p$ and $q$, respectively. We stress that, even if the features contained in $x_{tr}$ refer to the last 72 hours, the ARIMAX model has been left free to extend the endogenous signal's influence on the forecast up to 7 previous days, that is $p=7$. However, for all the considered locations, the grid search returned $p<=3$. 

\paragraph{\textbf{Random Forests and Quantile Random Forests}} \label{sec:RF}
The random forest (RF) algorithm independently fits several decision trees, each trained on different datasets, created from the original one through random re-sampling of the observations and keeping only a fraction of the overall features, chosen at random \citep{Hastie2009}. The final prediction of the RF is then a (possibly weighted) average of the trees' responses. One important variant of RF algorithms are Quantile Regression Forests (QRF); the main difference from RF is that QRF keeps the value of all the observations in the fitted trees' nodes, not just their mean, and assesses the conditional distribution based on this information. In this paper, we have used the Matlab \texttt{TreeBagger} class, which implements the QRF algorithm described in \cite{Meinhausen2014}.

\paragraph{\textbf{Tree-based boosting algorithms}}
Boosting algorithms employ additive training: starting from a constant model, at each iteration a new tree or any other so called "weak learner" $h_k(x)$ is added to the overall model $F_k(x)$, so that $F_{k+1}(x) = F_k(x) + \eta h_k(x)$ where $\eta \leq 1$ is a hyper-parameter denoting the learning rate, which helps reducing over-fitting.
The Least-squares gradient boosting (LSBoost) algorithm applies boosting in functional space: each weak learner $h$ tries to learn the gradient (with respect to the previous model $F_{k}(x)$) of the least-squares loss function. In other words, $h_k$ is fitted on the overall prediction error at iteration $k-1$. \\
A different approach is used by the XGBoost algorithm \citep{Chen2016}, which fits the additive model $F_k(x)$ in parameter space, that is, using a second-order approximation of the loss, as a function of the parameters of the weak learners (decision trees). This approximation and other techniques used by XGBoost (like approximate histogram search for selecting splitting points in the trees) result in a speedup of the training process, with respect to LSBoost or RF algorithms, without sacrificing accuracy. At the same time, the algorithm introduces quadratic penalization on the parameter's value and on the overall complexity of the trees, which parameters can be tuned to further mitigate over-fitting.\\ 
In addition to the QRF algorithm, in this paper we used a second algorithm that is able to assess the conditional probability distribution of the predictions: Natural Gradient Boosting (NGBoost) \citep{Duan2019}. While none of the previous algorithms introduced assumptions on the probability distribution of the observations, NGBoost explicitly fits the parameters of a parametric probability distribution on each observation. This is made possible by exploiting the tree structure of the underlying weak learner, since observations in the same leaves share the same probability distribution's parameters. The algorithm is fitted in functional space, but instead of directly learning the maximum likelihood gradient, the authors propose to correct it with the Fisher Information. This results in fitting the so-called natural gradient, which makes the learning process invariant to reparametrization of the underlying probability distribution. \\
We fitted LSBoost models using Matlab's \texttt{fitrensemble} function, tuning its hyper-parameters via Bayesian optimization and using 5 folds cross validation. The XGBoost and NGBoost algorithm were fitted using their official \texttt{xgboost} and \texttt{ngboost} python packages, respectively, while hyper-parameters were selected using grid search, always using a 5 folds cross validation strategy. We highlight that tuning the hyper-parameters in cross validation mitigates overfitting issues of the regression algorithms.

\section{Key performance indicators}  \label{sec:kpi}
The performance of the forecasting algorithms introduced in Section \ref{sec:algo} has been evaluated using to the following standard performance indicators:

\begin{linenomath*}
\begin{align}
\text{RMSE} &= \sqrt{\frac{1}{n}\sum_{i=1}^n(\hat{y}_i-y_i)^2}\\
\text{MAE} &= \frac{1}{n}\sum_{i=1}^n\left|\hat{y}_i-y_i \right|\\
\text{MAPE} &= \frac{100\%}{n}\sum_{i=1}^n\left|\frac{y_i-\hat{y}_i}{y_i} \right| \\
\text{S} &= 1-\frac{\text{RMSE}}{\text{RMSE}_{\text{pers}}}\\
\text{A} &= \frac{100\%}{n}\sum_{i=1}^{n}\mathbb{1}\left\{C(\hat{y}_i)=C(y_i) \right\}
\end{align}
\end{linenomath*}

where the acronyms, respectively, stand for Root Mean Squared Error, Mean Absolute Error, Mean Absolute Percentage Error, forecast Skill and Accuracy. 
$\text{RMSE}_{\text{pers}}$ is the RMSE of the persistence model, i.e. the model where the prediction at day $D+1$ is equal to the measured value at day $D$.
$C(y_i)$ is the function that associates every measured or forecasted value to the respective class, explicitly given by 
\begin{linenomath*}
\begin{equation}
\label{eq:classes}
C(y_i) = \left\{
\begin{aligned}
1 \quad &\text{ if }& 0 &< y_i \leq 60, \\
2 \quad &\text{ if }& 60 &< y_i \leq 120,\\
3 \quad &\text{ if }& 120 &< y_i \leq 135, \\
4 \quad &\text{ if }& 135 &< y_i \leq 180,\\
5 \quad &\text{ if }& 180 &< y_i \leq 240, \\
6 \quad &\text{ if }& 240 &< y_i.
\end{aligned}
\right.
\end{equation}
\end{linenomath*}

These values are the thresholds of classes of increasing severity of air pollution as indicated by the Swiss society of air protection officers (Cercl'Air) \citep{SchweizerischeGesellschaftderLufthygiene-Fachleute2019}. Class 3 is especially narrow compared to the other classes, as a result it will be harder for the regression forecasting algorithms to correctly predict this class.\\
Finally, we evaluated those algorithms which also returned conditional distributions, that is, QRF and NGBoost, using two additional KPIs. The first one is the reliability \citep{Pinson2010}, defined as
\begin{linenomath*}
\begin{equation}
    \text{R}(\tau)=\frac{1}{n}\sum_{i=1}^n\mathbb{1}\{y_i<\hat{y}_{i,\tau}\},
\end{equation}
\end{linenomath*}
where $\hat{y}_{i,\tau}$ is the quantile predicted by the algorithm at the level $\tau\in [0, 1]$. This KPI calculates how many of the total number of measured values are indeed lower than the quantile predicted on the same observations. If the forecasting algorithm were perfect, the R$(\tau)$ curve would lie on the bisector of the first quadrant.\\
The second probabilistic KPI is the average quantile loss function, also known as pinball loss \citep{Bentzien2014}:
\begin{linenomath*}
\begin{equation}
    \bar{\rho}(\tau)=\frac{1}{n}\sum_{i=1}^n \rho_{\tau}(y_i-\hat{y}_{i,\tau}),
\end{equation}
\end{linenomath*}
where the function $\rho_{\tau}(x)$ is defined as
\begin{linenomath*}
\begin{equation}
    \rho_{\tau}(x) = \left\{
    \begin{array}{lcl}
       \tau \left|x\right|  &\text{if}& x \geq 0,\\
        (1-\tau)\left|x\right|  &\text{if}& x < 0,
    \end{array}
    \right.
\end{equation}
\end{linenomath*}
This KPI measures how narrow is the predicted probability density function around the observations. It can be shown that this loss is minimized, independently of the underlying distribution which generated the data, when the predicted quantiles are the true ones. It should be noted that for $\tau=0.5$, the corresponding value $\bar{\rho}(\tau)$ is half the value of the MAE statistic. 
To further evaluate the performance of the quantiles as a single score, we also integrate $\bar{\rho}(\tau)$ over the $[0,1]$ interval, as outlined in \cite{Gneiting2007b}. Thus we define
\begin{linenomath*}
\begin{equation}
    \text{Q-score} = \int_0^1 \bar{\rho}(\tau)\, \mathrm{d}\tau.
\end{equation}
\end{linenomath*}

\section{Results}
\subsection{Feature number and feature selection methods comparison} \label{sec:NumFeat}
\begin{figure}
    \centering
    \includegraphics[width=\columnwidth]{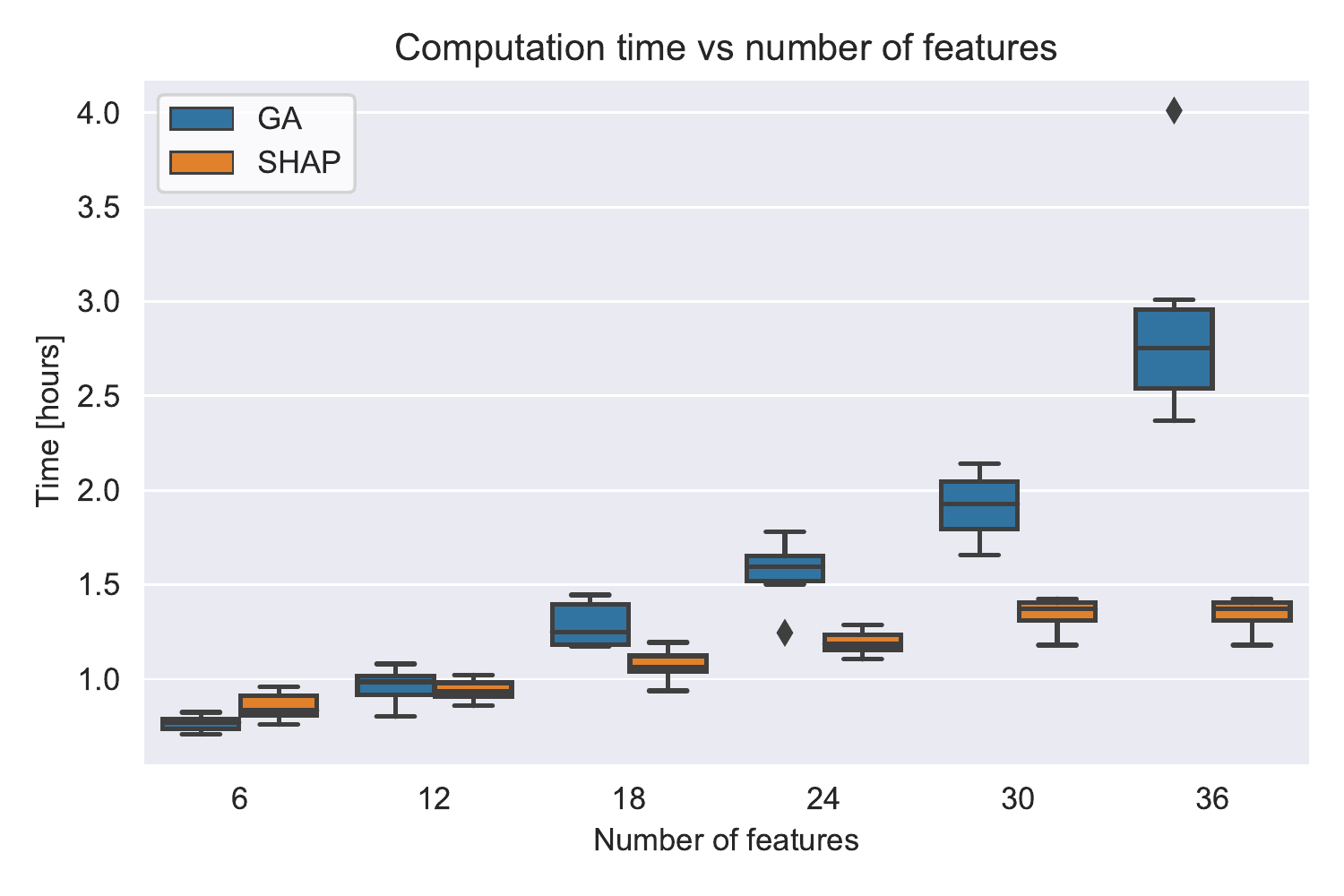}
    \caption{Boxplots for the computational time as a function of increasing number of features $k$ and feature selection method. Each boxplot contains values from the 7-folds CV, for three different locations.}
    \label{fig:timeVSfeat}
\end{figure}

To investigate the impact model complexity on the general quality of the results, we varied the number of features $k$ from 6 to 36 in steps of 6. Due to the relatively high computational time required to run this experiment, the models were calculated only for the 3 stations of Bioggio, Chiasso and Locarno, on the data from 2015 to 2018, using a 7 folds cross-validation. Fig. \ref{fig:timeVSfeat} shows the computational time, for all the models, as a function of the number of features and the feature selection method, for the three selected stations.

\begin{figure}
    \centering
    \includegraphics[width=\columnwidth]{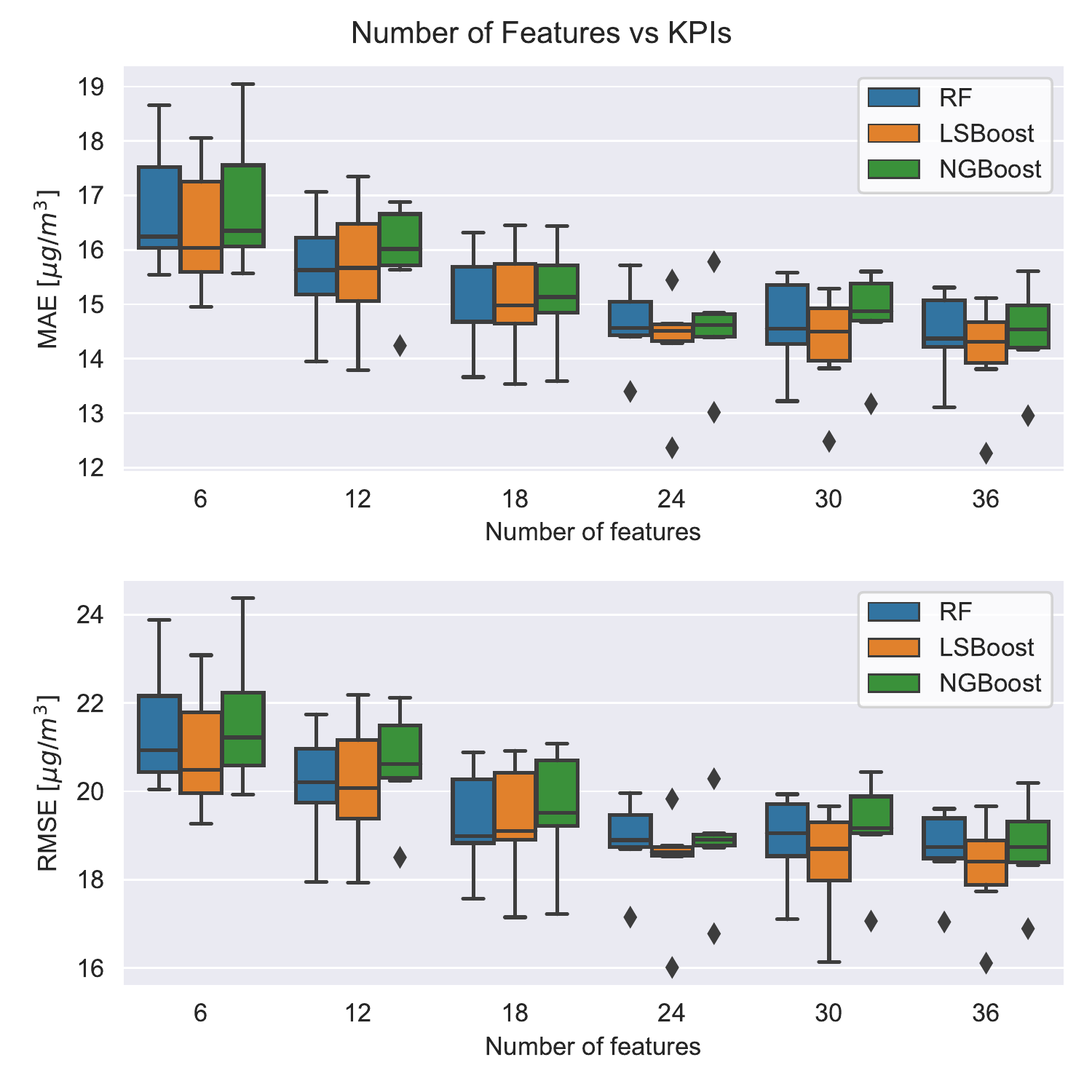}
    \caption{Boxplots for the MAE and RMSE, with respect to the number of features $k$. Each boxplot contains values from the 7-folds CV, for three different locations.}
    \label{fig:KPIVSfeat}
\end{figure}

The time needed to perform the feature selection and prediction of the results increases exponentially as a function of the number of features for the feature selection based on GA, while it shows a linear trend for the feature selection based on Shapley values. The CV computation could take up to 4 hours on our machine, a 16 core Intel i9-7960X CPU @ 2.80 GHz with 128 GB of RAM.\\
Similarly, using features selected with SHAP, we compared the performance of a subset of forecasting algorithms by gradually increasing the number of features (Fig. \ref{fig:KPIVSfeat}). The errors of the models do not appear to decrease after about 24 features. Consequently, we decided to construct our models for the main analysis with $k=30$ features. This number is a good compromise between the quality of the model and the computation time.

\begin{figure}
    \centering
    \includegraphics[width=1.0\columnwidth]{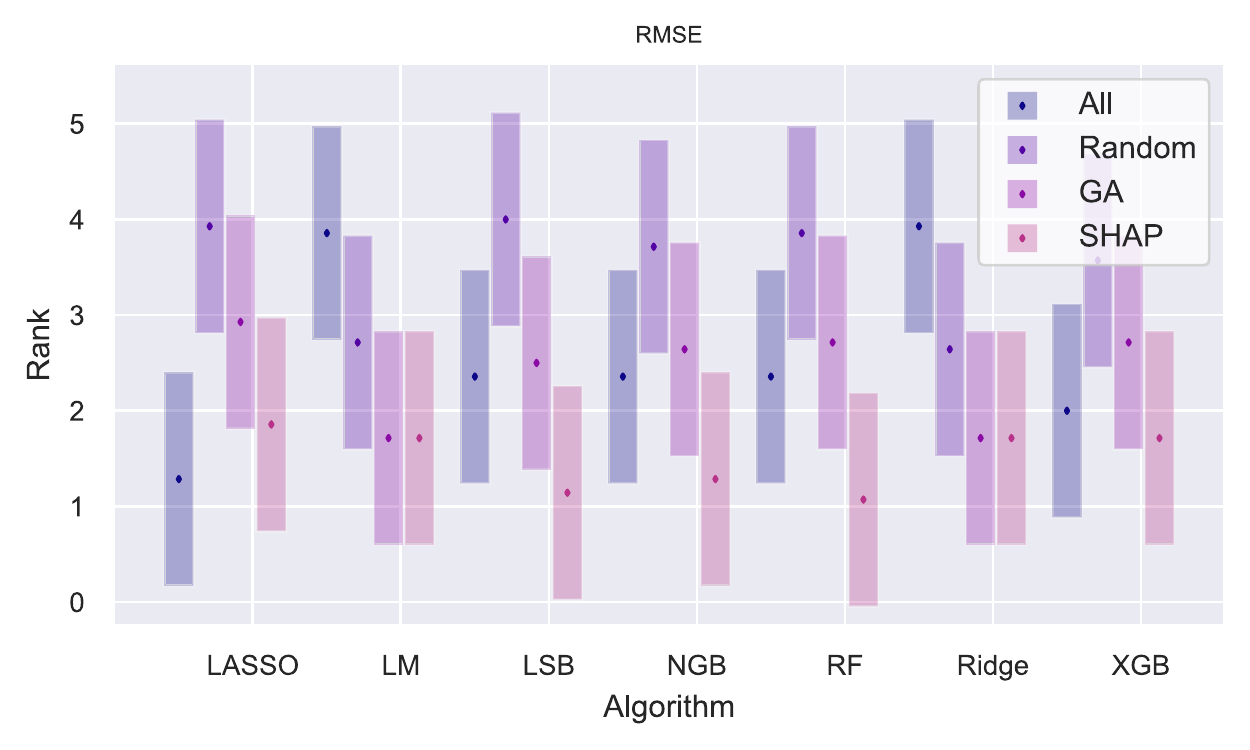}
    \caption{Nemenyi test for comparison of RMSE performance for the four feature selection methods across all locations}    
    \label{fig:nemenyi}
\end{figure}

To investigate whether feature selection effectively improves the quality of prediction and to determine which is the most efficient technique, we compared the performance of the algorithms across all locations using  four methods: GA, Shapley values (SHAP), random features and all available features (no feature selection). For the first three methods, 30 features are either selected according to the corresponding method or picked at random, while in the last case all the available features are used to build the model.\\
To compare the performances of the different models, we used Nemenyi statistical tests \citep{Hollander1999}, a post-hoc pairwise test, which is used to compare a set of $m$ different models on a group of $n$ independent experiments. Firstly, a matrix $R \in \mathrm{R}^{n\times m}$ whose elements $r_{i,j}$ are the ranks for experiment $i$ and model $j$, is obtained. Then, the mean rank for each model is retrieved through column-wise averages of $R$. The performance of the two models is identified as significantly different by the Nemenyi test if the corresponding average ranks differ by at least the critical difference
\begin{linenomath*}
\begin{equation}\label{eq:conf}
CD=q_{\alpha, m} \sqrt{\frac{m(m+1)}{12 n}}
\end{equation}
\end{linenomath*}
where $q_{\alpha}$ is the quantile $\alpha$ of the Studentized range statistic with $m$ samples, here setted to $\alpha=0.9$. We implemented the Nemenyi test in python following the implementation in the \texttt{tsutils} R package \citep{tsutils}.  The Nemenyi test is usually performed after a Friedman’s test, which is a non-parametric analog of variance for a randomized block design; this can be considered as non-parametric version of a one-way ANOVA with repeated measures. More details on the difference and implementation of the two tests can be found in \cite{Dale2006}.\\
The results of Nemeyi test can be observed in Fig. \ref{fig:nemenyi}. In Table \ref{tab:comparison}, we show the KPIs calculated for a few selected forecasting algorithms for the locations with best, average and worst RMSE statistics, namely Bioggio \textit{MOR}, Chiasso \textit{MOR}, and Mendrisio \textit{EVE}.

\begin{table*}[pos=htbp]
\centering
\setlength{\tabcolsep}{0.4em} 
{\renewcommand{\arraystretch}{1.2}
\caption{Comparison between different feature selections for the best, average and worst stations}
\label{tab:comparison}
\hspace*{-0.2cm}
\begin{tabular}{|c|l|l|l|l|l|l|l|l|l|l|l|l|l|}
\hline
\multicolumn{2}{|l|}{}                   & \multicolumn{4}{c|}{\textbf{BIO MOR}}                                                                                         & \multicolumn{4}{c|}{\textbf{CHI MOR}}                                                                                         & \multicolumn{4}{c|}{\textbf{MEN EVE}}                                                                                         \\ \cline{3-14} 
\multicolumn{2}{|l|}{\multirow{-2}{*}{}} & \multicolumn{1}{c|}{SHAP}     & \multicolumn{1}{c|}{GA}       & \multicolumn{1}{c|}{All}      & \multicolumn{1}{c|}{Rand}     & \multicolumn{1}{c|}{SHAP}     & \multicolumn{1}{c|}{GA}       & \multicolumn{1}{c|}{All}      & \multicolumn{1}{c|}{Rand}     & \multicolumn{1}{c|}{SHAP}     & \multicolumn{1}{c|}{GA}       & \multicolumn{1}{c|}{All}      & \multicolumn{1}{c|}{Rand}     \\ \hline
                            & MAE        & 13.46 & 14.17 & 14.41 & 15.58 & 14.50 & 15.18 & 15.27 & 19.02 & 16.69 & 17.24 & 16.90 & 19.84 \\
                            & RMSE       & 17.68 & 18.38 & 18.98 & 20.35 & 19.36 & 21.25 & 20.03 & 24.21 & 21.59 & 22.61 & 21.70 & 24.92 \\
                            & MAPE       & 10.34 & 11.06 & 11.29 & 12.32 & 10.65 & 11.05 & 11.33 & 14.42 & 12.75 & 13.57 & 13.52 & 15.59 \\
                            & S          & 0.430 & 0.408 & 0.388 & 0.344 & 0.419 & 0.362 & 0.399 & 0.273 & 0.333 & 0.302 & 0.330 & 0.231 \\
\multirow{-5}{*}{LSBoost}   & Accuracy   & 71.43 & 68.71 & 71.43 & 69.39 & 71.43 & 71.43 & 70.75 & 60.54 & 65.07 & 60.96 & 59.59 & 59.59 \\ \hline
                            & MAE        & 14.05 & 14.32 & 14.63 & 15.55 & 15.84 & 15.66 & 17.34 & 21.13 & 16.78 & 17.69 & 17.54 & 21.43 \\
                            & RMSE       & 18.58 & 19.01 & 19.44 & 21.09 & 20.87 & 21.26 & 21.58 & 26.27 & 22.23 & 22.80 & 22.27 & 26.94 \\
                            & MAPE       & 10.91 & 11.17 & 11.67 & 12.24 & 11.77 & 11.54 & 12.95 & 16.56 & 13.05 & 13.66 & 14.01 & 16.88 \\
                            & S          & 0.401 & 0.387 & 0.373 & 0.320 & 0.374 & 0.362 & 0.352 & 0.212 & 0.314 & 0.296 & 0.313 & 0.169 \\
\multirow{-5}{*}{XGBoost}   & Accuracy   & 68.71 & 72.79 & 65.99 & 70.07 & 70.07 & 70.07 & 69.39 & 57.14 & 65.75 & 62.33 & 60.27 & 56.85 \\ \hline
                            & MAE        & 13.85 & 14.38 & 15.02 & 15.64 & 15.04 & 15.88 & 15.85 & 18.72 & 16.12 & 17.05 & 17.54 & 20.14 \\
                            & RMSE       & 18.50 & 18.70 & 20.04 & 20.60 & 19.90 & 21.64 & 21.11 & 23.58 & 21.32 & 22.79 & 22.76 & 25.46 \\
                            & MAPE       & 10.73 & 11.13 & 11.96 & 12.36 & 11.06 & 11.85 & 11.98 & 14.26 & 12.54 & 13.29 & 14.12 & 16.04 \\
                            & S          & 0.404 & 0.397 & 0.354 & 0.336 & 0.403 & 0.351 & 0.367 & 0.292 & 0.342 & 0.297 & 0.298 & 0.214 \\
\multirow{-5}{*}{NGBoost}   & Accuracy   & 68.71 & 68.71 & 68.71 & 68.03 & 69.39 & 70.07 & 67.35 & 59.86 & 66.44 & 60.27 & 58.90 & 57.53 \\ \hline
\end{tabular}}
\end{table*}

The results show that the Shapley values approach obtains the best rank for all the ensemble methods across all the locations. For Ridge and LM, we essentially have a tie between SHAP and GA. In the case of LASSO, the approach that uses all features obtains the best rank. This is likely explained by the fact that LASSO, by design, inherently performs feature selection. We can conclude that performing a preliminary feature selection using Shapley values analysis is preferable in terms of KPIs quality and computation time. The SHAP feature selection is faster compared to GA and All and avoids needlessly slowing down the predictive algorithms.

\subsection{Feature importance and interactions} \label{sec:feat_imp}
To assess the importance of each feature in predicting the maximum ozone concentration, we applied once again the \texttt{shap} library introduced in Section \ref{sec:SHAP}, using NGBoost algorithm as the $f(x_{tr},\Theta)$ regression model. We thus analyzed the most relevant features for the 7 locations and their interactions. Table \ref{tab:most_relevant} summarizes the 3 most relevant features for the considered locations. For 5 of the considered cases, the most important feature was a value of forecasted temperature for both the \textit{MOR} and \textit{EVE} predictions. For the other cases, the most important feature was a past value of ozone concentration. The three most relevant features include forecasted values of the temperature, locally or at a nearby station, and past measured ozone concentration for all the locations.

\begin{table}[]
	\caption{First 3 most relevant features for the \textit{MOR} and \textit{EVE} predictions for the 7 considered locations. Colors refer to the two dominant classes of variables. Green: NWP forecasted temperaturesk, blue: measured ozone.}
	\label{tab:most_relevant}
	\setlength{\tabcolsep}{0.5em} 
    {\renewcommand{\arraystretch}{1.6}
	\begin{tabular}{|c|c|l|l|l|}
\hline
\multicolumn{2}{|c|}{\textbf{Location}} & \multicolumn{1}{c|}{\textbf{f1}}                          & \multicolumn{1}{c|}{\textbf{f2}}                       & \multicolumn{1}{c|}{\textbf{f3}}                     \\ \hline
                            & MOR       & \cellcolor[HTML]{D4EFDF}$\widehat{T}^{p_4}_{b_2,mean}$    & $NO2^{l_3}_{m_{0}}$                                    & \cellcolor[HTML]{D4EFDF}$\widehat{T}^{p_4}_{s_{15}}$ \\ \cline{2-5} 
\multirow{-2}{*}{BIO}       & EVE       & \cellcolor[HTML]{D4EFDF}$\widehat{T}^{p_4}_{s_{29}}$      & \cellcolor[HTML]{D4EFDF}$\widehat{T}^{p_4}_{b_4,mean}$ & \cellcolor[HTML]{D6EAF8}$O3^{l_3}_{m_{0}}$           \\ \hline
                            & MOR       & \cellcolor[HTML]{D4EFDF}$\widehat{T}^{p_6}_{chunk2,mean}$ & \cellcolor[HTML]{D6EAF8}$O3^{l_5}_{m_{11}}$            & \cellcolor[HTML]{D4EFDF}$\widehat{T}^{p_5}_{s_{13}}$ \\ \cline{2-5} 
\multirow{-2}{*}{CHI}       & EVE       & \cellcolor[HTML]{D4EFDF}$\widehat{T}^{p_6}_{b_4,mean}$    & \cellcolor[HTML]{D6EAF8}$O3^{l_5}_{m_{1}}$             & \cellcolor[HTML]{D4EFDF}$\widehat{T}^{p_5}_{s_{27}}$ \\ \hline
                            & MOR       & \cellcolor[HTML]{D4EFDF}$\widehat{T}^{p_6}_{b_2,mean}$    & \cellcolor[HTML]{D4EFDF}$\widehat{T}^{p_6}_{s_{13}}$   & \cellcolor[HTML]{FFFFFF}$NO2^{l_5}_{m_{0}}$          \\ \cline{2-5} 
\multirow{-2}{*}{MEN}       & EVE       & \cellcolor[HTML]{D4EFDF}$\widehat{T}^{p_6}_{b_4,mean}$    & \cellcolor[HTML]{D6EAF8}$O3^{l_5}_{m_{1}}$             & \cellcolor[HTML]{FFFFFF}$\widehat{G}^{p_5}_{PM}$           \\ \hline
                            & MOR       & \cellcolor[HTML]{D4EFDF}$\widehat{T}^{p_2}_{b_2,mean}$    & \cellcolor[HTML]{FFFFFF}$\widehat{G}^{p_2}_{b_1,mean}$        & \cellcolor[HTML]{D4EFDF}$\widehat{T}^{p_2}_{s_{11}}$ \\ \cline{2-5} 
\multirow{-2}{*}{LOC}       & EVE       & \cellcolor[HTML]{D4EFDF}$\widehat{T}^{p_2}_{b_4,max}$     & \cellcolor[HTML]{D6EAF8}$O3^{l_1}_{m_{0}}$             & \cellcolor[HTML]{D4EFDF}$\widehat{T}^{p_1}_{s_{23}}$ \\ \hline
                            & MOR       & \cellcolor[HTML]{D4EFDF}$\widehat{T}^{p_2}_{s_{9}}$       & \cellcolor[HTML]{D4EFDF}$\widehat{T}^{p_2}_{b_2,mean}$ & \cellcolor[HTML]{D4EFDF}$\widehat{T}^{p_1}_{s_{8}}$  \\ \cline{2-5} 
\multirow{-2}{*}{BRI}       & EVE       & \cellcolor[HTML]{D4EFDF}$\widehat{T}^{p_2}_{b_4,max}$     & \cellcolor[HTML]{D4EFDF}$\widehat{T}^{p_1}_{s_{23}}$   & \cellcolor[HTML]{D4EFDF}$\widehat{T}^{p_1}_{s_{25}}$ \\ \hline
                            & MOR       & \cellcolor[HTML]{D4EFDF}$\widehat{T}^{p_2}_{b_4,max}$     & \cellcolor[HTML]{D6EAF8}$O3^{l_7}_{m_{13}}$            & \cellcolor[HTML]{FFFFFF}$YO3^{l_7}$             \\ \cline{2-5} 
\multirow{-2}{*}{SAG}       & EVE       & \cellcolor[HTML]{D6EAF8}$O3^{l_7}_{m_{1}}$                & \cellcolor[HTML]{D6EAF8}$O3^{l_7}_{m_{0}}$             & \cellcolor[HTML]{D6EAF8}$O3^{l_7}_{24h}$             \\ \hline
                            & MOR       & \cellcolor[HTML]{D6EAF8}$O3^{l_4}_{24h,mean}$             & \cellcolor[HTML]{D4EFDF}$\widehat{T}^{p_4}_{s_{16}}$   & $Wd^{p_4}_{s_{3}}$                               \\ \cline{2-5} 
\multirow{-2}{*}{TES}       & EVE       & \cellcolor[HTML]{D6EAF8}$O3^{l_4}_{m_{0}}$                & \cellcolor[HTML]{D4EFDF}$\widehat{T}^{p_4}_{s_{29}}$   & \cellcolor[HTML]{D4EFDF}$\widehat{T}^{p_4}_{s_{30}}$ \\ \hline
\end{tabular}}
\end{table}

\begin{figure}
    \centering
    \includegraphics[width=0.47\textwidth]{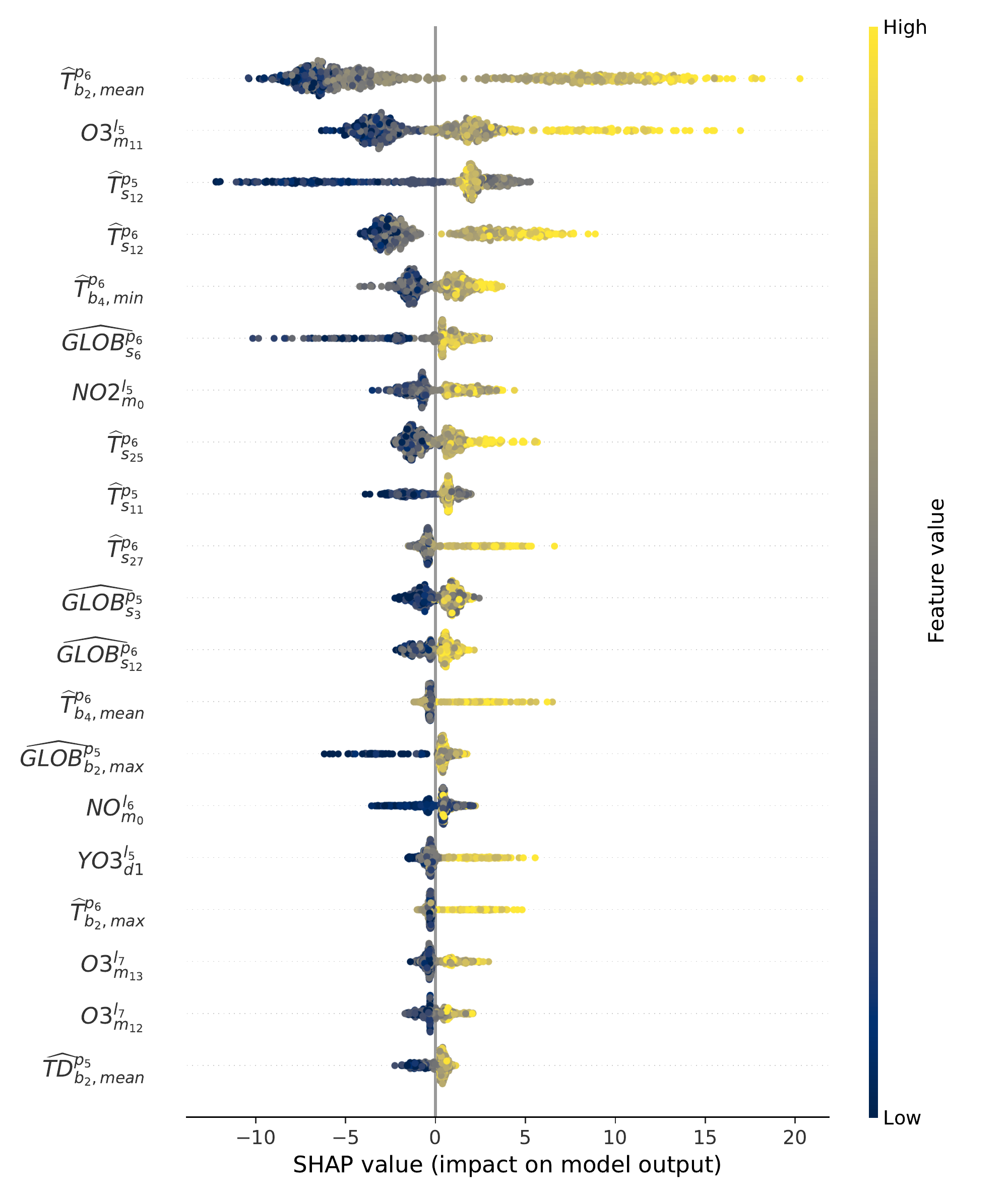}
    \caption{Feature importance for each observation, for Chiasso \textit{MOR}.}    
    \label{fig:features_importance}
\end{figure}

\begin{figure}
	\centering
	\includegraphics[width=1\linewidth]{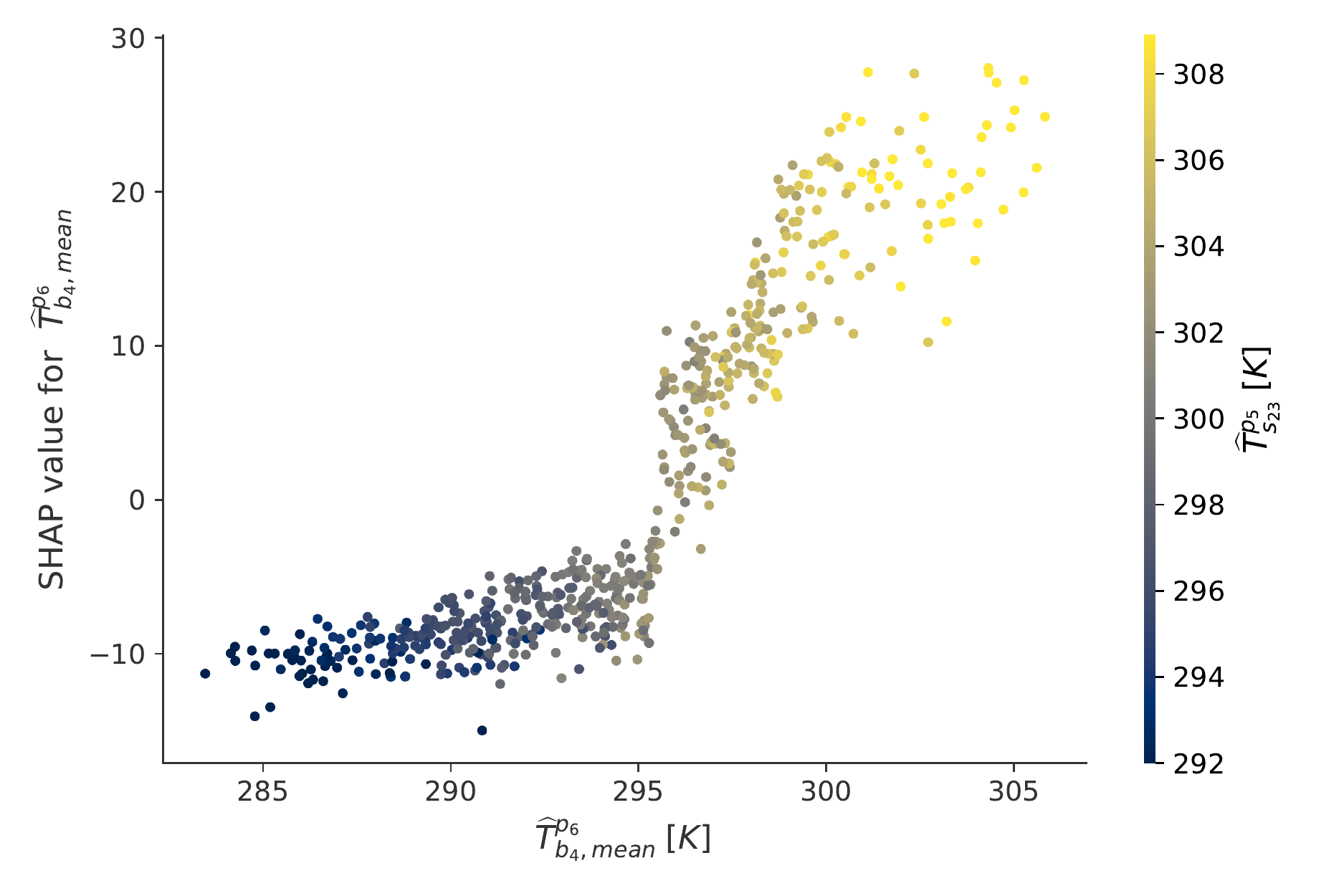}
	\caption{Feature importance interaction between mean of forecasted temperatures in Sagno between 14:00 and 23:00, and forecasted temperature in Chiasso at 17:00, on the \textit{EVE} prediction of ozone in Chiasso.}
	\label{fig:temperature_dep}
\end{figure}

\begin{figure}
	\centering
	\includegraphics[width=0.9\linewidth]{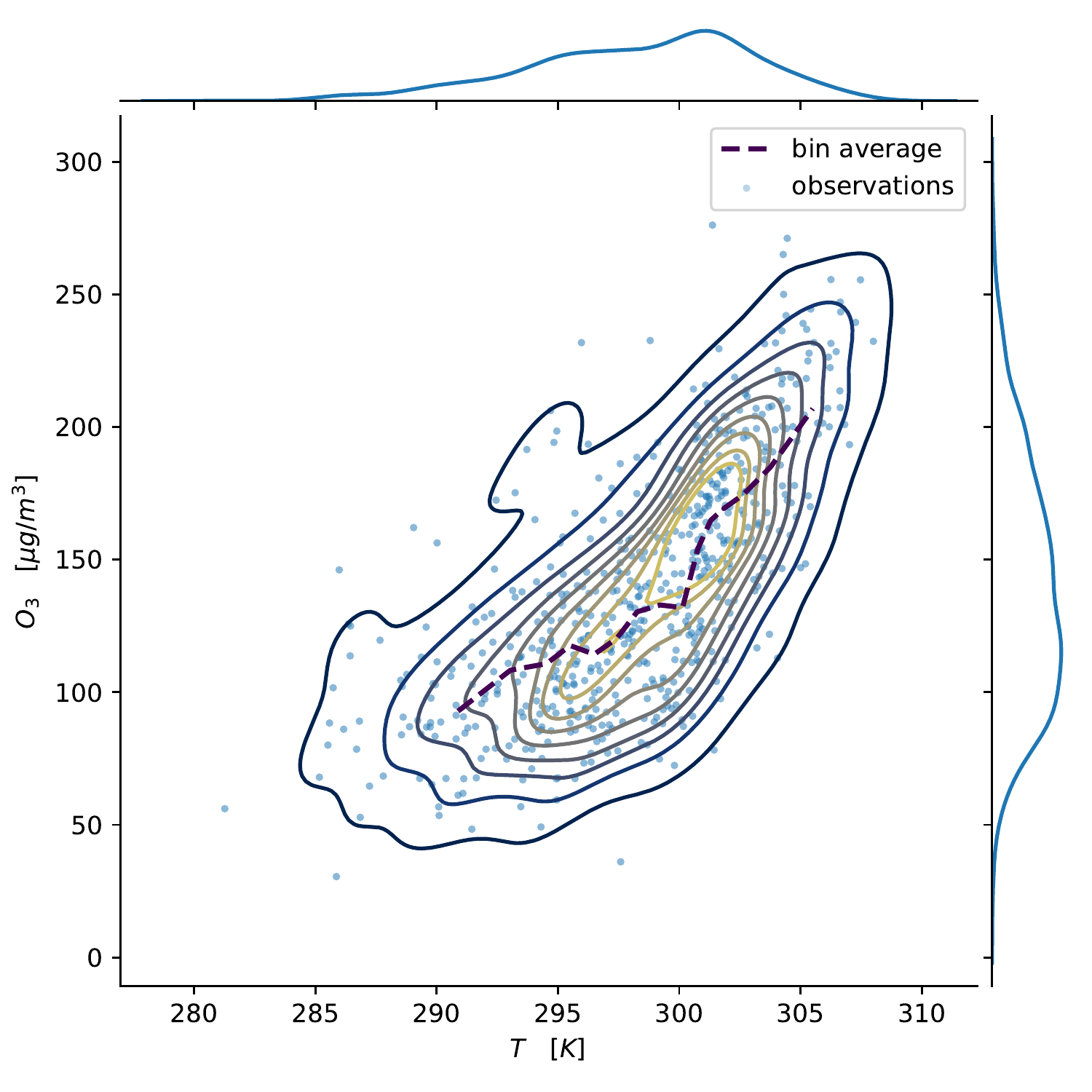}
	\caption{Partial dependence plot between instantaneously measured $T$ and \ce{O_3} in Chiasso. Continuous lines are iso-density surfaces estimated with a kernel density estimator.}
	\label{fig:temperature_dep_raw}
\end{figure}

\begin{figure} 
	\centering 
	\includegraphics[width=\linewidth]{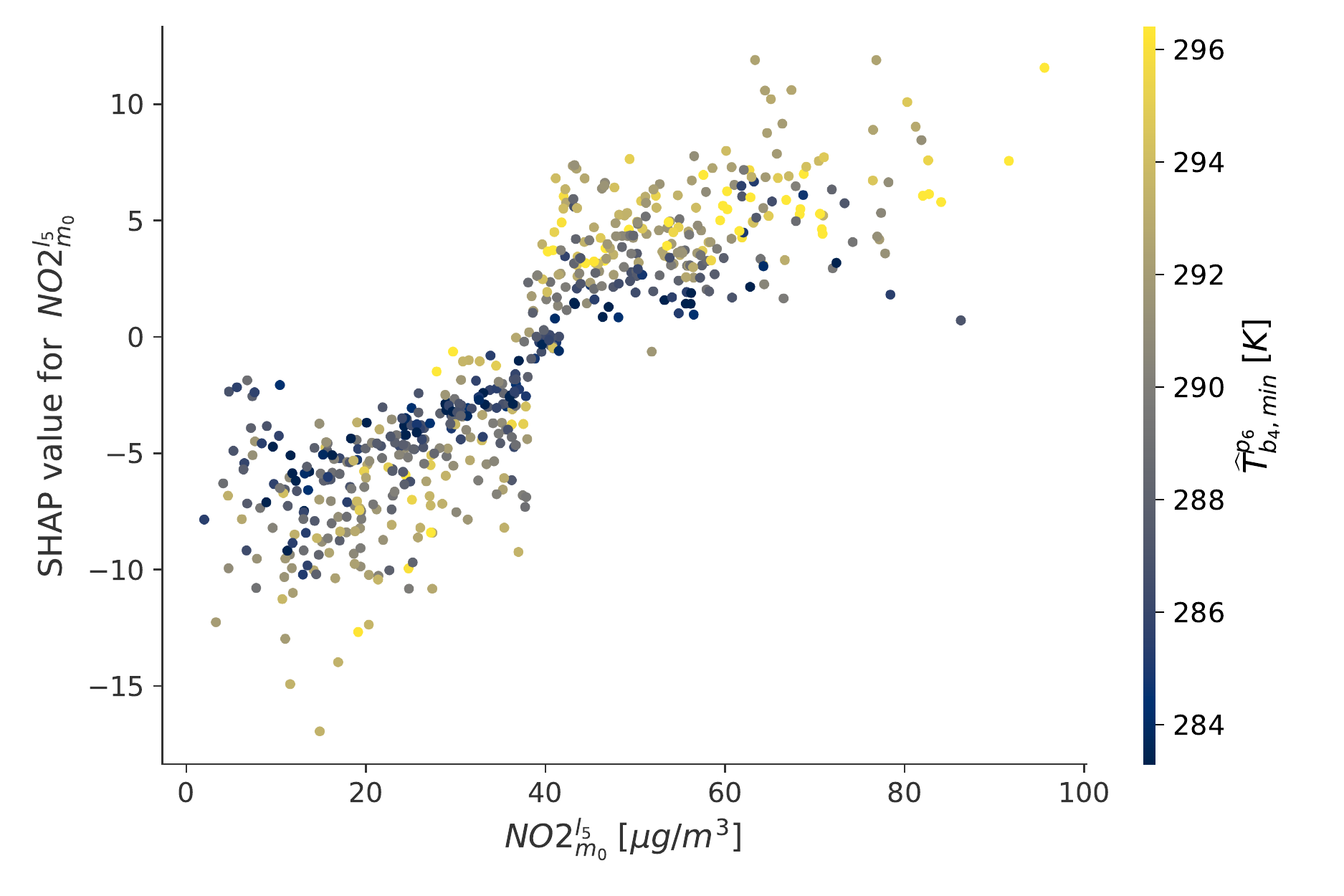}
	\caption{Feature importance interaction between measured $\ce{NO_2}$ in Chiasso and temperature forecasted in Sagno, on the \textit{MOR} prediction of ozone in Mendrisio.}    
	\label{fig:no2_1}
\end{figure}

\begin{figure} 
	\centering 
	\includegraphics[width=\linewidth]{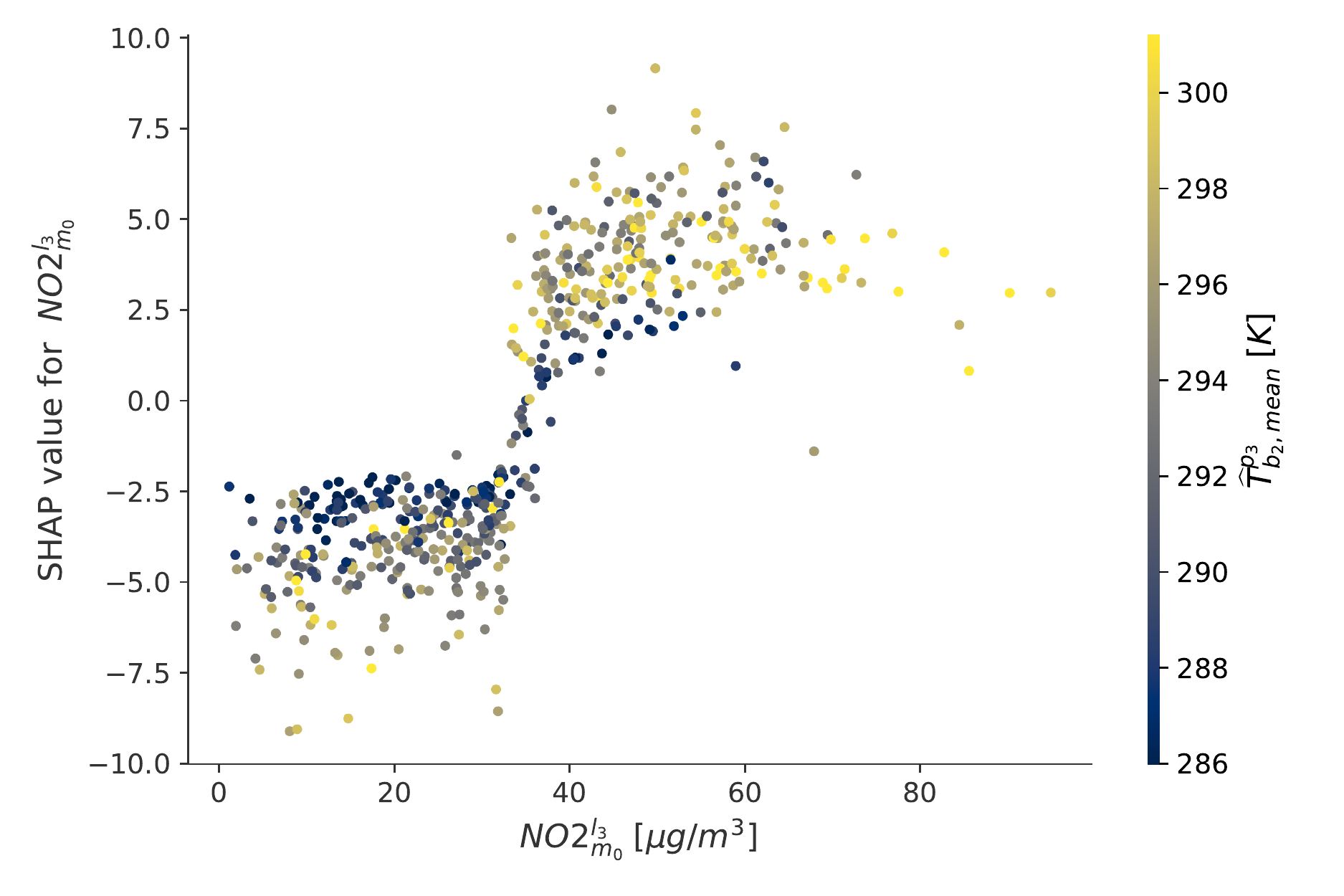}
	\caption{Feature importance interaction between measured $\ce{NO_2}$ in Bioggio and temperature forecasted in Sagno, on the \textit{MOR} prediction of ozone in Bioggio.}
	\label{fig:no2_2}
\end{figure}

\subsubsection{Temperature influence}
In Fig. \ref{fig:features_importance}, we show the influence of single observations on the ozone forecast, divided by feature, for the location of Chiasso. The most relevant predictor is $\widehat{T}^{p_6}_{b_4,mean}$, which is the forecasted temperature at a nearby location. Several chemical reactions and pathways influence the chemistry of ozone formation and removal in the troposphere \citep{Lu2019,Crutzen2016,Monks2015}, such that imputing temperature's influence to just one of them is hardly satisfactory, and other factors and processes must be considered. The photochemical processes are dominant factors in determining the ozone peak concentrations, together with thermal and radiative conditions. Other more physical processes, such as local wind systems and diurnal boundary layer dynamics, are of secondary importance in our study, since it focuses on the generally well-mixed diurnal boundary layer.  We aim to forecast the peak concentrations under well-mixed afternoon conditions rather than the ozone concentration variations within the diurnal cycle. We stress that some phenomena, like the role of the effective height of the boundary layer for the air mass characteristics (precursor mixture, photochemical reactivity), determining the volume available to dilute the emissions, and the role of large-scale advection, are not directly considered by the forecasting algorithms, and that our analysis is limited to the variables selected by the feature selection procedure. However, since the temperature is causally correlated with the height of the boundary layer, we're indirectly accounting for some of its effect by using temperature as a predictor. Fig. \ref{fig:temperature_dep} shows the Shapley values for the mean value of the temperatures forecasted in Sagno between 14:00 and 23:00, and the interaction with the mean forecasted temperature in Chiasso at 17:00. We can observe how the influence of the forecasted temperature on the maximum \ce{O3} peak does not follow the typical exponential form of a chemical rate constant, but rather a sigmoid. A similar sigmoid-like functional relation between these two variables can be observed in all the other locations. This functional form can be explained explicitly modeling the main chemical reactions involving the formation and destruction of \ce{O3}, and their temperature-dependent rate coefficients.  Authors in \citep{Pusede2015, Pusede2014} modeled the interdependence of \ce{O3} production rate, \ce{NO_x} and temperature. Initially, the authors modeled the main chemical tropospheric \ce{O3} formation processes. Then, they replaced the volatile organic compounds (VOC) reactivity with a functional relation learned from observations between the latter and daily maximum temperature. At last, they found a  sigmoid-like influence between the daily maximum temperature on \ce{O3} production, for concentrations of \ce{NO_x} greater than 6 ppb. A similar sigmoid-like dependence of \ce{O3} formation rates and the temperature was found in \cite{Walcek1995}, both considering or disregarding a linear correlation between maximum daily temperature and solar irradiance (Fig. 4a and Fig. 5 in the reference, respectively). In all these studies, the authors kept the influence of other variables, such as VOCs concentration and irradiance fixed, or treated them as parameters. This is equivalent to performing a sensitivity analysis on the rate of change of \ce{O3}, which is exactly the scope of the Shapley variables. In this sense, the findings of the aforementioned authors are compatible with Fig. \ref{fig:temperature_dep}.
We stress that the sigmoid-like importance of temperature in predicting \ce{O3} concentration cannot be directly extracted from the raw data. To better explain this, we show in Fig. \ref{fig:temperature_dep_raw} the partial dependence plot of the measured $O_3$, that is, the target variable and the measured temperature. As we can see, raw data show an exponential-like relation, very similar to what was found in \cite{Walcek1995}, Fig. 1a, where measurements of maximum hourly \ce{O3} concentration and temperature from the New Jersey urban region are plotted.

\subsubsection{$\boldmath{\ce{NO_2}}$ importance interactions}
Solar irradiance and \ce{NO_x} concentration play a major role in the photochemistry of tropospheric ozone \citep{Crutzen2016,Walcek1995}. In Fig. \ref{fig:no2_1} and \ref{fig:no2_2}, the influence of the measured \ce{NO_2} and its interaction with forecasted temperature on the \textit{MOR} prediction of the \ce{O3} in Bioggio and Sagno, is shown. We can see for both cases how the sigmoid-like \ce{NO_2} importance becomes more pronounced for increasing forecasted temperatures. This means that at high concentrations of \ce{NO_2}, higher temperatures accelerate \ce{O3} formation rate. On the other hand, Fig. \ref{fig:no2_1} and \ref{fig:no2_2} suggest that at low concentration of \ce{NO_2}, \ce{O3} generation from \ce{NO_2} becomes increasingly important with increasing temperature, with respect to other \ce{O3} formation concurring processes. This results in low \ce{NO_2} concentrations being more significant predictors of low \ce{O3} concentrations at high ambient temperatures. These conclusions cannot be explained with the same kind of analysis carried out for example in \cite{Pusede2015}, where only the \ce{O3} formation rate with respect to \ce{NO_x} and temperature is investigated, and not its relative importance over other concurrent photochemical pathways.   

\subsection{Performance of the regression models}\label{sec:results}
In Table \ref{tab:results} we present the results of our study with respect to the KPIs introduced in Section \ref{sec:kpi}, obtained by first applying the SHAP feature selection approach as described in \ref{sec:SHAP}.

\begin{figure*}
\centering
\includegraphics[width=0.95\textwidth]{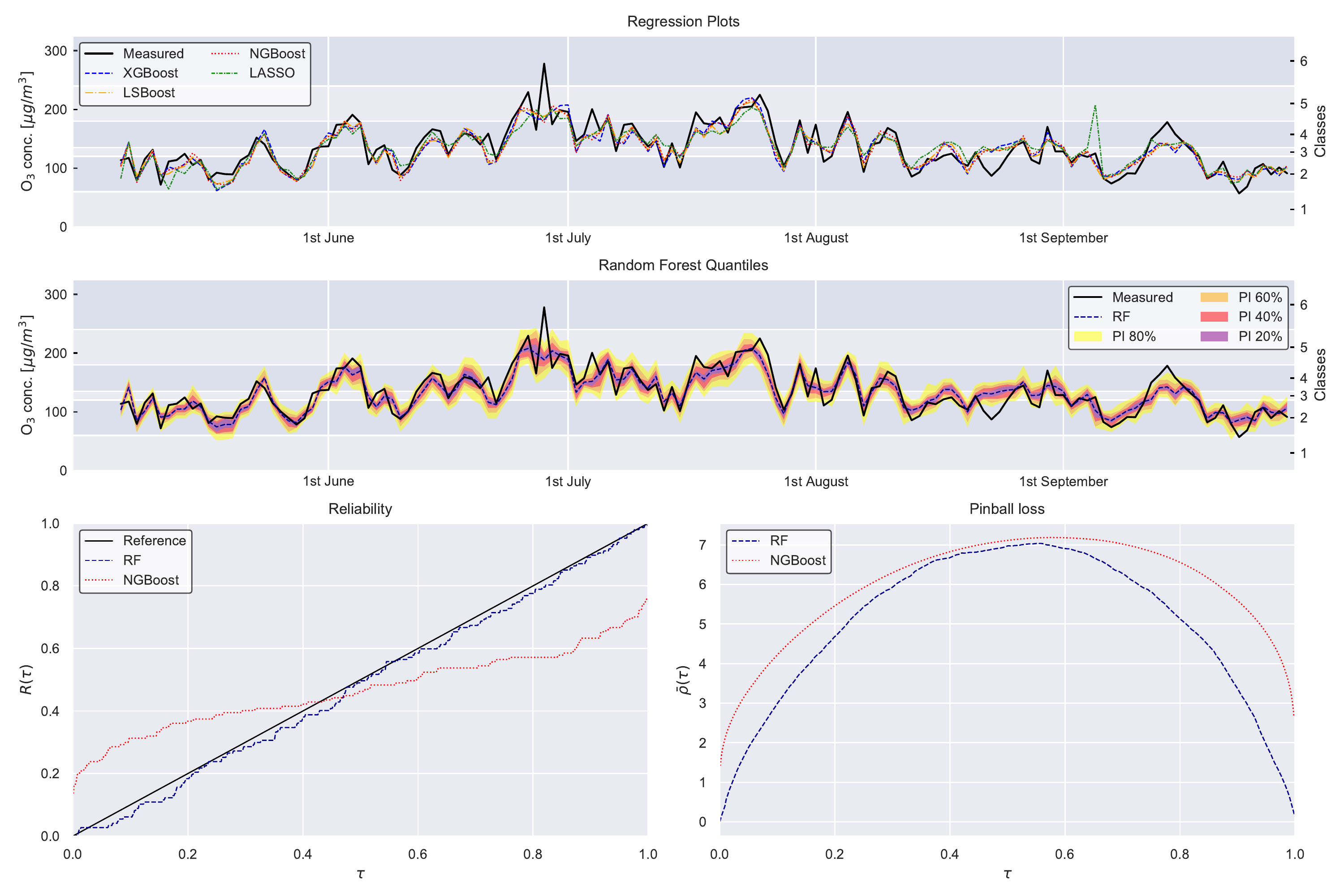}
\caption{Result plots of the station yielding the best results, Bioggio MOR.}    
\label{fig:res_bio_mor}
\end{figure*}
\begin{figure*}
\centering
\includegraphics[width=0.95\textwidth]{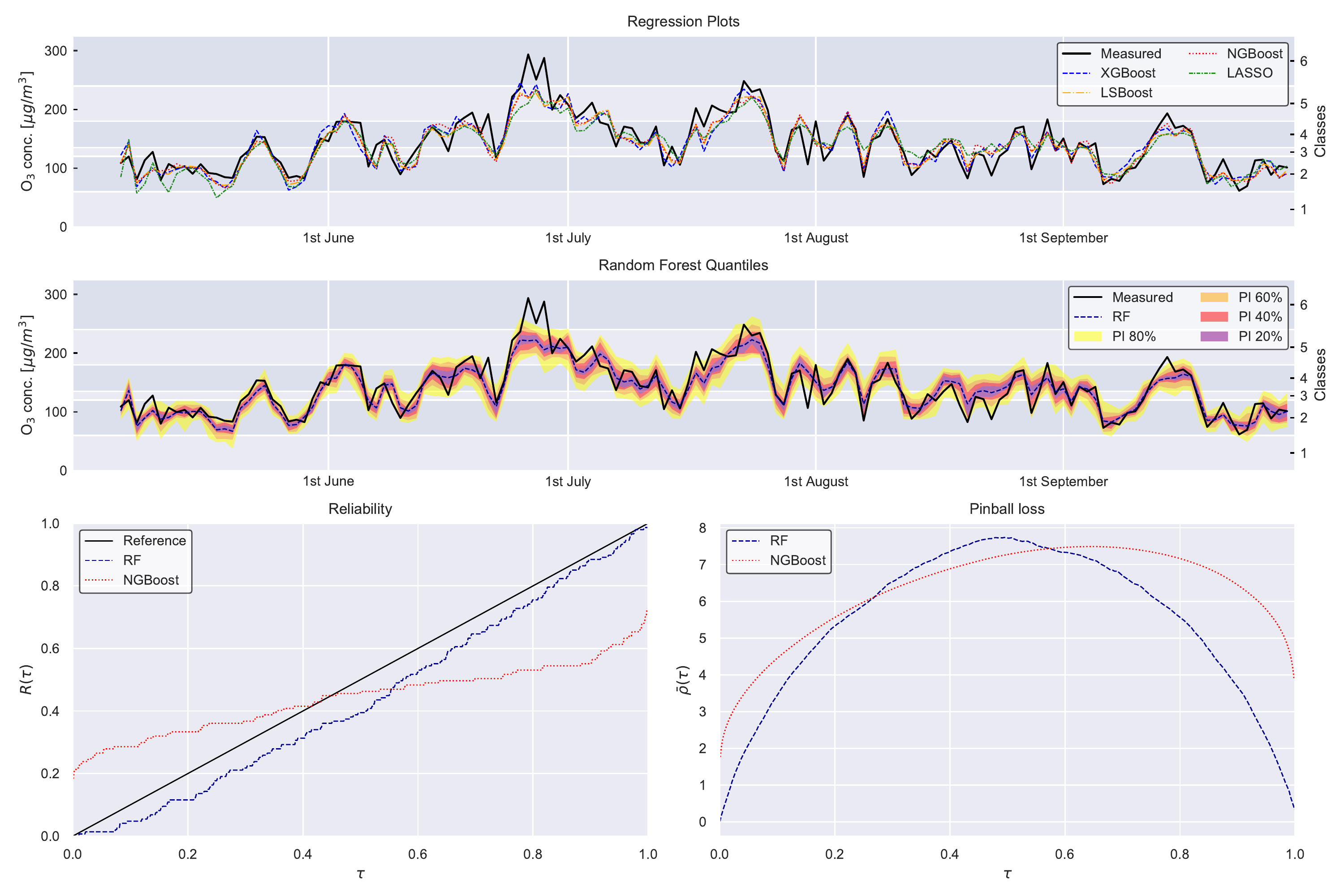}
\caption{Result plots of a station yielding average results, Chiasso MOR.}    
\label{fig:res_chi_mor}
\end{figure*}
\begin{figure*}
\centering
\includegraphics[width=0.95\textwidth]{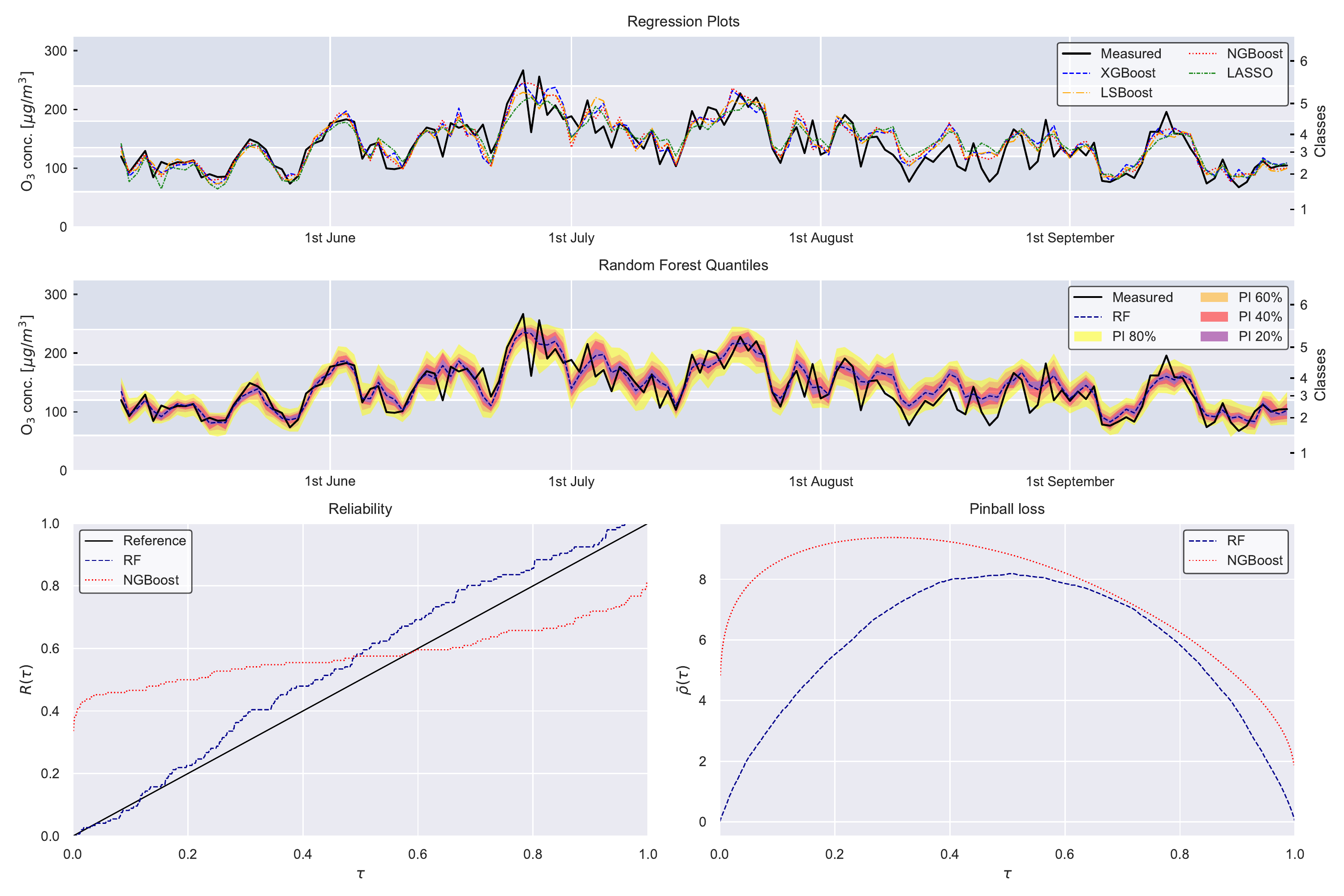}
\caption{Result plots of the station yielding the worst results, Mendrisio EVE.}    
\label{fig:res_men_eve}
\end{figure*}

\begin{table*}
\centering
\setlength{\tabcolsep}{0.4em} 
{\renewcommand{\arraystretch}{1.2}
\caption{Main results of the study with Shapley values feature selection. Boldface indicate the lowest RMSE for the corresponding location}
\label{tab:results}
\hspace*{-0.2cm}
\begin{tabular}{|c|l|rr|rr|rr|rr|rr|rr|rr|}
\hline
\multicolumn{2}{|l|}{\multirow{2}{*}{}}  & \multicolumn{2}{c|}{\textbf{BIO}}                            & \multicolumn{2}{c|}{\textbf{CHI}}                            & \multicolumn{2}{c|}{\textbf{MEN}}                            & \multicolumn{2}{c|}{\textbf{LOC}}                            & \multicolumn{2}{c|}{\textbf{BRI}}                            & \multicolumn{2}{c|}{\textbf{SAG}}                            & \multicolumn{2}{c|}{\textbf{TES}}                            \\ \cline{3-16} 
\multicolumn{2}{|l|}{}   & \multicolumn{1}{c|}{MOR} & \multicolumn{1}{c|}{EVE} & \multicolumn{1}{c|}{MOR} & \multicolumn{1}{c|}{EVE} & \multicolumn{1}{c|}{MOR} & \multicolumn{1}{c|}{EVE} & \multicolumn{1}{c|}{MOR} & \multicolumn{1}{c|}{EVE} & \multicolumn{1}{c|}{MOR} & \multicolumn{1}{c|}{EVE} & \multicolumn{1}{c|}{MOR} & \multicolumn{1}{c|}{EVE} & \multicolumn{1}{c|}{MOR} & \multicolumn{1}{c|}{EVE} \\ \hline
\multirow{6}{*}{RF}      & MAE      & 13.81                                 & 15.28            & 15.20                                 & 15.82            & 15.70                                 & 16.10            & 14.08                                 & 14.43            & 14.52            & 14.71            & 14.88            & 15.37            & 13.64                                 & 15.03            \\
                         & RMSE     & 18.07                                 & 19.82            & 20.21                                 & 21.66            & 20.54                                 & \textbf{21.30}   & \textbf{18.33}                        & 19.01            & 19.03            & \textbf{20.20}   & 19.48            & 20.62            & 19.41                                 & 21.63            \\
                         & MAPE     & 10.74                                 & 11.83            & 11.13                                 & 11.69            & 12.35                                 & 12.65            & 11.78                                 & 12.36            & 12.49            & 12.80            & 11.93            & 12.25            & 12.08                                 & 13.82            \\
                         & S        & 0.418                                 & 0.364            & 0.394                                 & 0.352            & 0.364                                 & 0.342            & 0.394                                 & 0.374            & 0.404            & 0.370            & 0.295            & 0.256            & 0.201                                 & 0.114            \\
                         & Accuracy & 69.39                                 & 67.81            & 66.67                                 & 65.75            & 65.99                                 & 62.33            & 68.03                                 & 62.33            & 66.67            & 67.12            & 61.76            & 62.22            & 66.33                                 & 65.98            \\
                         & Q-score  & 4.987                                 & 5.396            & 5.452                                 & 5.765            & 5.590                                 & 5.749            & 5.133                                 & 5.239            & 5.208            & 5.364            & 5.259            & 5.662            & 4.494                                 & 4.961            \\ \hline
\multirow{5}{*}{LSBoost} & MAE      & 13.46                                 & 13.96            & 14.50                                 & 15.56            & 14.66                                 & 16.69            & 14.20                                 & 14.30            & 14.25            & 14.50            & 15.08            & 14.59            & 13.37                                 & 14.86            \\
                         & RMSE     & \textbf{17.68}                        & \textbf{18.18}   & \textbf{19.36}                        & \textbf{21.03}   & \textbf{19.50}                        & 21.59            & 18.58                                 & 19.03            & \textbf{18.78}   & 20.53            & \textbf{19.39}   & \textbf{19.73}   & 18.89                                 & 20.73            \\
                         & MAPE     & 10.34                                 & 10.78            & 10.65                                 & 11.47            & 11.54                                 & 12.75            & 11.71                                 & 11.98            & 12.03            & 12.53            & 12.09            & 11.53            & 11.71                                 & 13.44            \\
                         & S        & 0.430                                 & 0.416            & 0.419                                 & 0.371            & 0.396                                 & 0.333            & 0.386                                 & 0.373            & 0.412            & 0.360            & 0.298            & 0.288            & 0.222                                 & 0.151            \\
                         & Accuracy & 71.43                                 & 67.81            & 71.43                                 & 67.12            & 66.67                                 & 65.07            & 67.35                                 & 69.18            & 68.03            & 69.18            & 61.76            & 61.48            & 70.41                                 & 64.95            \\ \hline
\multirow{5}{*}{XGBoost} & MAE      & 14.05                                 & 15.10            & 15.84                                 & 16.80            & 16.01                                 & 16.78            & 14.81                                 & 15.84            & 15.57            & 15.18            & 15.92            & 16.96            & 13.73                                 & 15.31            \\
                         & RMSE     & 18.58                                 & 19.58            & 20.87                                 & 22.04            & 21.45                                 & 22.23            & 18.85                                 & 22.21            & 19.58            & 21.31            & 20.25            & 22.44            & 19.33                                 & 21.27            \\
                         & MAPE     & 10.91                                 & 11.73            & 11.77                                 & 12.30            & 12.35                                 & 13.05            & 12.17                                 & 13.83            & 13.17            & 13.19            & 12.66            & 13.73            & 11.92                                 & 13.77            \\
                         & S        & 0.401                                 & 0.371            & 0.374                                 & 0.341            & 0.336                                 & 0.314            & 0.377                                 & 0.268            & 0.387            & 0.335            & 0.266            & 0.190            & 0.204                                 & 0.129            \\
                         & Accuracy & 68.71                                 & 73.29            & 70.07                                 & 67.12            & 65.31                                 & 65.75            & 66.67                                 & 70.55            & 68.03            & 65.07            & 61.76            & 56.30            & 65.31                                 & 64.95            \\ \hline
\multirow{6}{*}{NGBoost} & MAE      & 13.85                                 & 14.83            & 15.04                                 & 16.23            & 14.84                                 & 16.12            & 14.96                                 & 14.02            & 15.18            & 15.09            & 19.57            & 19.77            & 19.34                                 & 21.76            \\
                         & RMSE     & 18.50                                 & 19.56            & 19.90                                 & 21.27            & 19.62                                 & 21.32            & 18.92                                 & \textbf{18.84}   & 19.78            & 20.37            & 25.36            & 26.44            & 26.12                                 & 28.92            \\
                         & MAPE     & 10.73                                 & 11.32            & 11.06                                 & 11.94            & 11.68                                 & 12.54            & 12.42                                 & 12.06            & 12.83            & 13.08            & 16.93            & 16.86            & 18.37                                 & 20.90            \\
                         & S        & 0.404                                 & 0.372            & 0.403                                 & 0.364            & 0.393                                 & 0.342            & 0.375                                 & 0.379            & 0.381            & 0.365            & 0.082            & 0.046            & -0.07                                 & -0.18            \\
                         & Accuracy & 68.71                                 & 69.18            & 69.39                                 & 65.75            & 65.99                                 & 66.44            & 61.90                                 & 67.12            & 65.31            & 69.18            & 47.79            & 50.37            & 59.18                                 & 52.58            \\
                         & Q-score  & 5.947                                 & 6.298            & 6.298                                 & 7.485            & 6.429                                 & 7.671            & 6.453                                 & 6.333            & 6.241            & 5.979            & 12.14            & 12.50            & 25.94                                 & 27.36            \\ \hline
\multirow{5}{*}{LM}      & MAE      & 15.27                                 & 15.64            & 17.46                                 & 17.49            & 16.55                                 & 26.42            & 16.12                                 & 27.26            & 15.32            & 73.43            & 15.14            & 14.67            & 13.01                                 & 171.7            \\
                         & RMSE     & 21.67                                 & 20.32            & 23.03                                 & 23.37            & 21.74                                 & 109.9            & 20.84                                 & 144.2            & 20.41            & 693.6            & 19.78            & 19.99            & 19.03                                 & 945.3            \\
                         & MAPE     & 12.01                                 & 12.32            & 13.38                                 & 13.62            & 12.92                                 & 21.12            & 13.67                                 & 22.76            & 12.95            & 62.56            & 12.22            & 11.60            & 11.25                                 & 154.0            \\
                         & S        & 0.301                                 & 0.347            & 0.309                                 & 0.301            & 0.327                                 & -2.39            & 0.311                                 & -3.75            & 0.361            & -20.6            & 0.284            & 0.279            & 0.217                                 & -37.7            \\
                         & Accuracy & 68.71                                 & 65.75            & 66.67                                 & 61.64            & 65.31                                 & 61.64            & 65.31                                 & 65.07            & 68.71            & 66.44            & 63.24            & 64.44            & 67.35                                 & 17.53            \\ \hline
\multirow{5}{*}{LASSO}   & MAE      & 15.76                                 & 15.91            & 17.54                                 & 17.68            & 17.94                                 & 17.56            & 15.33                                 & 16.26            & 15.38            & 18.11            & 15.59            & 14.87            & 12.95                                 & 14.18            \\
                         & RMSE     & 21.13                                 & 20.43            & 22.99                                 & 23.58            & 23.22                                 & 22.89            & 20.18                                 & 21.07            & 20.73            & 23.82            & 20.11            & 20.32            & \textbf{18.83}                        & \textbf{20.64}   \\
                         & MAPE     & 12.30                                 & 12.40            & 13.31                                 & 13.45            & 14.12                                 & 14.00            & 12.99                                 & 13.87            & 13.27            & 15.79            & 12.52            & 11.67            & 11.35                                 & 12.92            \\
                         & S        & 0.319                                 & 0.344            & 0.310                                 & 0.295            & 0.281                                 & 0.293            & 0.333                                 & 0.306            & 0.351            & 0.257            & 0.272            & 0.267            & 0.225                                 & 0.155            \\
                         & Accuracy & 65.99                                 & 64.38            & 63.27                                 & 61.64            & 58.50                                 & 59.59            & 65.31                                 & 63.70            & 68.03            & 62.33            & 58.82            & 63.70            & 65.31                                 & 64.95            \\ \hline
\multirow{5}{*}{Ridge}   & MAE      & 15.23                                 & 15.63            & 17.46                                 & 17.47            & 16.51                                 & 26.42            & 16.12                                 & 27.26            & 15.16            & 71.56            & 15.13            & 14.67            & 12.97                                 & 168.5            \\
                         & RMSE     & 21.61                                 & 20.32            & 23.03                                 & 23.32            & 21.70                                 & 109.9            & 20.84                                 & 144.2            & 20.23            & 670.5            & 19.77            & 19.99            & 19.00                                 & 914.6            \\
                         & MAPE     & 11.97                                 & 12.31            & 13.38                                 & 13.59            & 12.88                                 & 21.12            & 13.67                                 & 22.76            & 12.83            & 60.96            & 12.22            & 11.61            & 11.21                                 & 151.1            \\
                         & S        & 0.304                                 & 0.347            & 0.309                                 & 0.302            & 0.328                                 & -2.39            & 0.311                                 & -3.75            & 0.367            & -19.9            & 0.284            & 0.279            & 0.218                                 & -36.4            \\
                         & Accuracy & 68.71                                 & 66.44            & 66.67                                 & 61.64            & 65.31                                 & 61.64            & 65.31                                 & 65.07            & 69.39            & 65.75            & 64.71            & 64.44            & 67.35                                 & 16.49            \\ \hline
\multirow{5}{*}{ARIMAX}  & MAE      & 17.57                                 & 16.87            & 17.12                                 & 17.72            & 16.57                                 & 18.05            & 14.8                                  & 14.74            & 16.3             & 14.94            & 16.25            & 14.38            & 12.89                                 & 13.32            \\
                         & RMSE     & 23.34                                 & 22.28            & 23.21                                 & 22.77            & 21.95                                 & 23.42            & 20.39                                 & 19.79            & 22.76            & 19.73            & 21.35            & 19.07            & 19.17                                 & 20.88            \\
                         & MAPE     & 13.42                                 & 13.05            & 12.92                                 & 13.26            & 12.44                                 & 13.94            & 12.33                                 & 12.43            & 13.41            & 12.81            & 13.09            & 11.44            & 10.89                                 & 11.6             \\
                         & S        & 0.25                                  & 0.28             & 0.3                                   & 0.32             & 0.32                                  & 0.28             & 0.32                                  & 0.35             & 0.29             & 0.38             & 0.21             & 0.29             & 0.71                                  & 0.69             \\
                         & Accuracy & 60.54                                 & 63.7             & 63.95                                 & 61.38            & 63.95                                 & 57.24            & 63.95                                 & 68.49            & 65.31            & 69.18            & 56.46            & 63.7             & 53.06                                 & 58.9             \\ \hline
\end{tabular}}
\end{table*}

In general, the two algorithms that overall perform best are LSBoost and NGBoost, since most of the lowest MAE and RMSE values and the highest accuracy values are concentrated in these two algorithms. In the station of Tesserete nearly one third of the total ozone measurements of summer 2019 were not available, so the results are not directly comparable with those of the other stations.

Table \ref{tab:results} shows that, for the same algorithm, the \textit{MOR} results are generally better than the \textit{EVE} counterpart, although the difference in some cases is slight. This is not surprising since the \textit{MOR} predictions also use the data gathered overnight, which are not available to the \textit{EVE} forecasters. However, there is an exception in Brione, where the MAE statistic is slightly higher at \textit{MOR} for all the ensemble algorithms. The best results are obtained in Bioggio, where both \textit{MOR} and \textit{EVE} forecasts have the best KPIs across all the stations.\\
The relatively small differences between \textit{MOR} and \textit{EVE} seem to indicate that the data gathered during the night are not particularly important. In fact, as Table \ref{tab:most_relevant} shows, the three most important features for each model essentially revolve around ozone measured during the peak of the previous day and forecasted temperature. In all models, a lot of importance is given to the forecasts of the upcoming afternoon and the measured ozone values of the previous afternoon. Arguably, the NWP forecasts are more precise in the morning, thus leading to more precise ozone predictions as well.\\
Fig. \ref{fig:res_bio_mor}, \ref{fig:res_chi_mor} and \ref{fig:res_men_eve} graphically illustrate the results for the best (Bioggio \textit{MOR}), average (Chiasso \textit{MOR}) and worst (Mendrisio \textit{EVE}) cases. Each figure is composed of 4 plots. The top one shows a comparison between the main forecasting algorithms and the measured values. The second plot shows the prediction intervals at levels 20\%, 40\%, 60\% and 80\% issued by the RF algorithm as well as the RF prediction. The third and fourth plots further investigate the goodness of fit of the quantiles of the RF and NGBoost algorithms.

\subsection{High peaks prediction}\label{subsec:high_peaks}
The analysis presented so far focused on the dataset in its entirety, aiming to provide the best KPIs over all the data, independently from the air pollution severity class. However, when predicting ozone concentrations, it is generally more important to be able to correctly predict high concentrations, \FloatBarrier \noindent as they can pose a health risk.\\
For this reason, we decided to prompt the predictor to focus more on high concentrations, in our case classes 4, 5 and 6 as defined in the equation \ref{eq:classes}, by introducing weighted training. 
We assigned different weights to the observations depending on the severity class they were in. We found that it was beneficial to assign a weight $w_1\in [20,200]$ to observations in class 6, $w_2\in [20,200]$ to observations in class 5, $w_3\in [10,20]$ to observations in class 4 and finally a fixed weight of 1 to all the other observations. The key idea is to find the optimal set of weights ${w_1, w_2, w_3}$ for each location and case that improves the forecasting quality at high concentrations. The same set of weights is used during feature selection with SHAP and NGBoost and to train the prediction algorithm with the selected features. Applying weights during feature selection should help select the most important features to recognize the highest ozone concentrations.\\
We sought to increase the classification accuracy of three particular subsets of our data using weighted training, by evaluating the best combination of weights to optimize prediction accuracy for: 
\begin{itemize}
    \item Values in classes 4, 5 and 6 (\ce{O_3} > $135 \mu g/m^3$)
    \item Values in classes 5 and 6 (\ce{O_3} > $180 \mu g/m^3$)
    \item Values in class 6 (\ce{O_3} > $240 \mu g/m^3$)
\end{itemize}

\begin{table}
\centering
\setlength{\tabcolsep}{0.4em} 
{\renewcommand{\arraystretch}{1.2}
\caption{Number of particular events registered in the analyzed stations in 2019}
\label{tab:events}
\begin{tabular}{|l|c|c|c|}
\hline
\multicolumn{1}{|c|}{\textbf{Station}} & \multicolumn{1}{c|}{\textbf{\vtop{\hbox{\strut Values above}\hbox{\strut 135  $\boldmath{\boldsymbol{\mu} g/m^3}$}\hbox{\strut cl. $\mathbf{\left[4,5,6\right]}$ }}}}  & \multicolumn{1}{c|}{\textbf{\vtop{\hbox{\strut Values above}\hbox{\strut 180 $\boldmath{\boldsymbol{\mu} g/m^3}$}\hbox{\strut cl. $\mathbf{\left[5,6\right]}$}}}} & \multicolumn{1}{c|}{\textbf{\vtop{\hbox{\strut Values above}\hbox{\strut 240 $\boldmath{\boldsymbol{\mu} g/m^3}$}\hbox{\strut cl. 6}}}} \\ \hline
Chiasso                                & 46                                             & 23                                             & 4                                              \\ \hline
Bioggio                                & 47                                             & 17                                             & 1                                              \\ \hline
Mendrisio                              & 70                                             & 22                                             & 2                                              \\ \hline
Locarno                                & 41                                             & 9                                              & 1                                              \\ \hline
\end{tabular}}
\end{table}

\begin{figure*}
\centering
\begin{subfigure}{0.32\textwidth}
  \includegraphics[width=\textwidth]{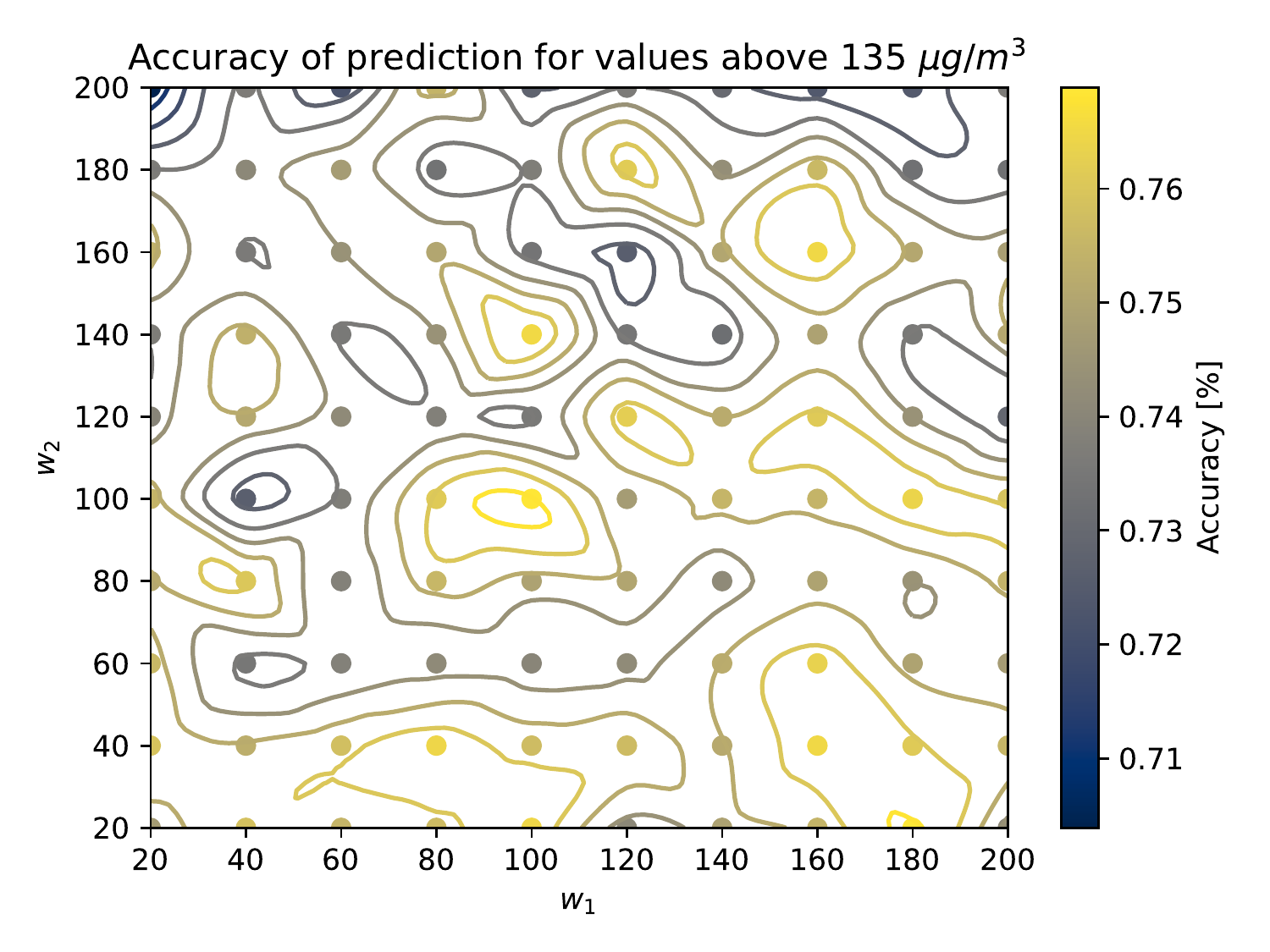}
  \caption{Optimization of classes [4, 5, 6]\\$w_3=10$} 
  \label{fig:grid_search_135_w3_1}
  \par\medskip 
  \includegraphics[width=\textwidth]{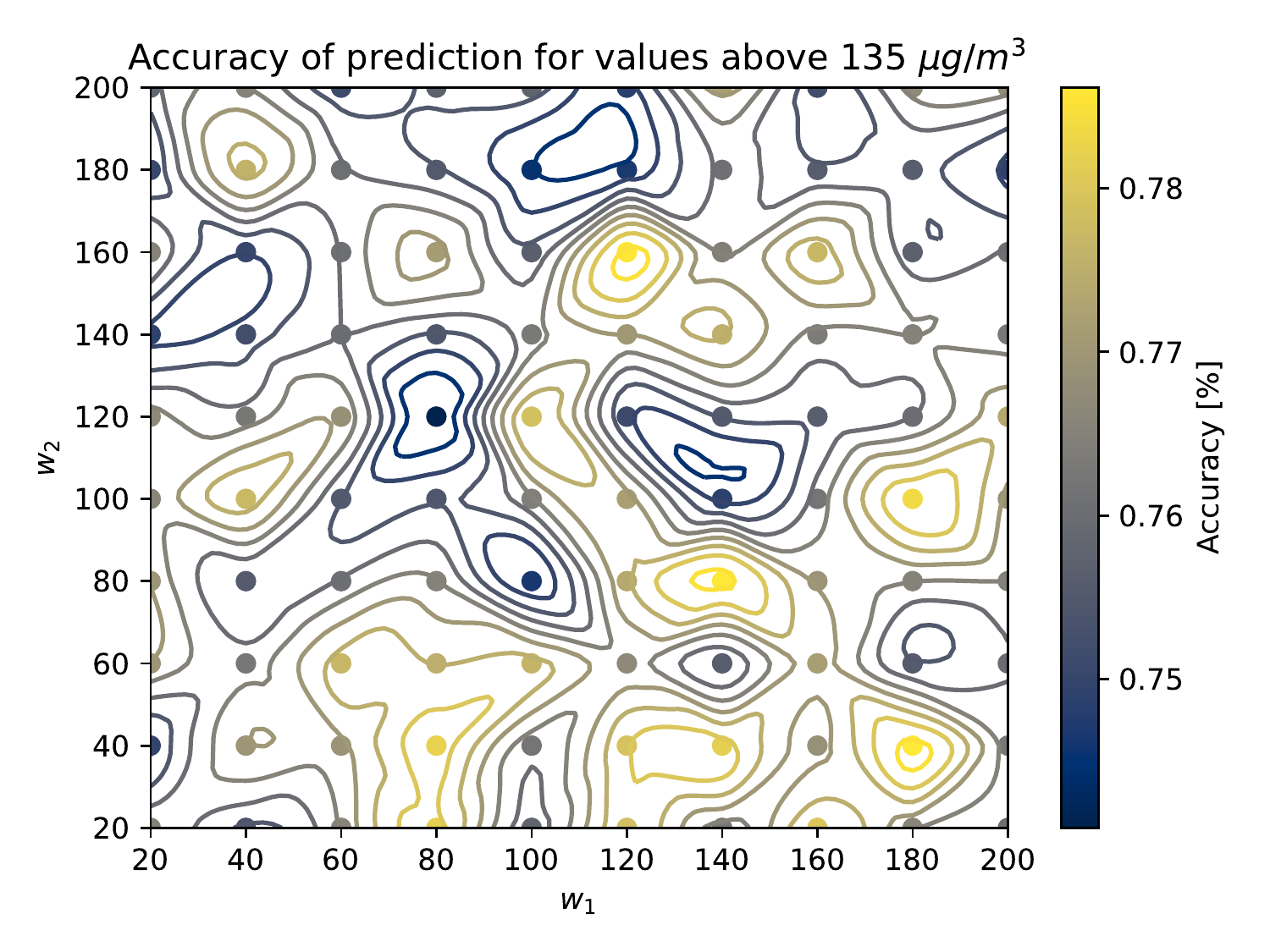}
  \caption{Optimization of classes [4, 5, 6]\\$w_3=20$}
  \label{fig:grid_search_135_w3_2}
\end{subfigure}
\hspace*{\fill}
\begin{subfigure}{0.32\textwidth}
  \includegraphics[width=\textwidth]{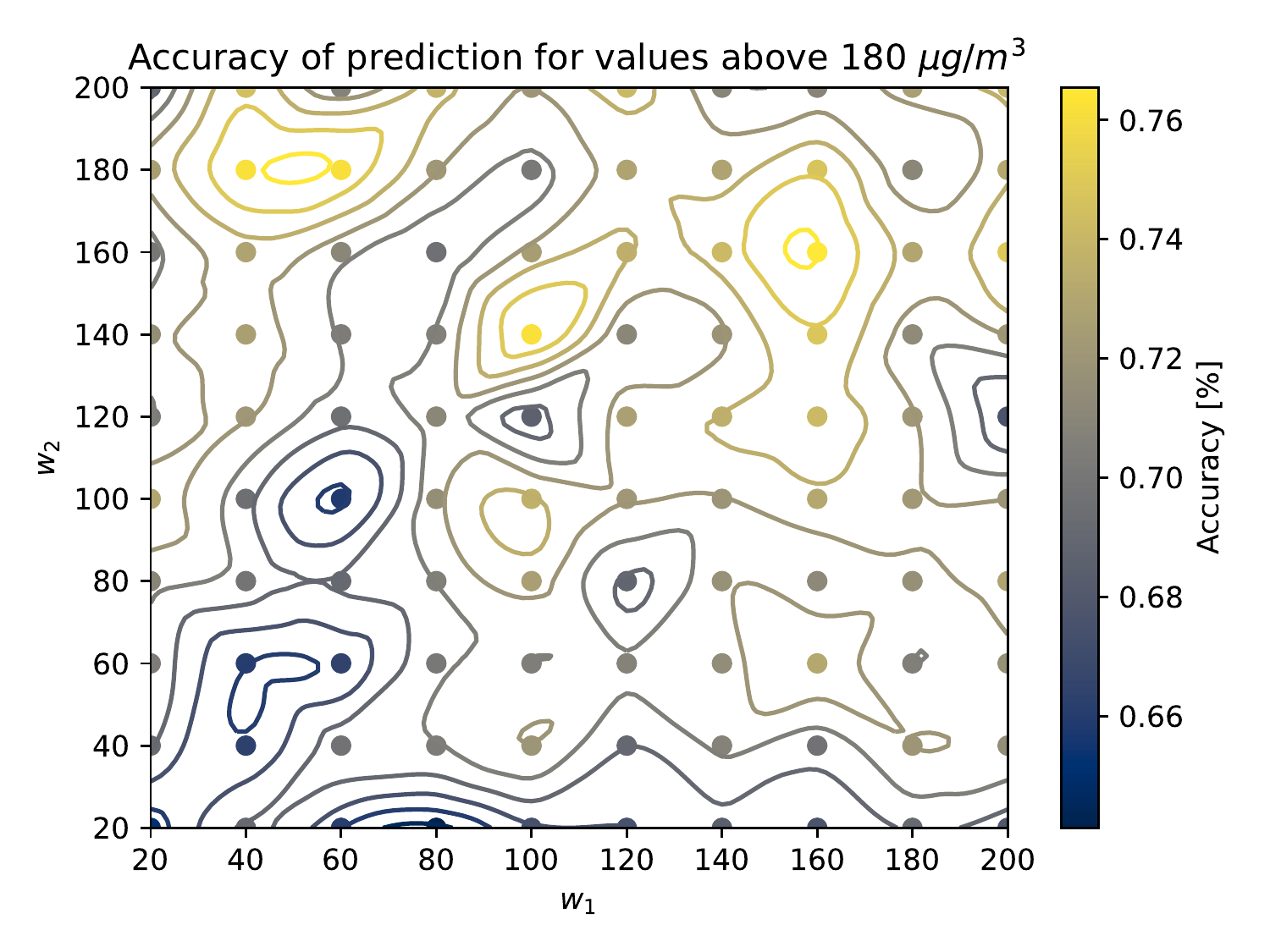}
  \caption{Optimization of classes [5, 6]\\$w_3=10$} 
  \label{fig:grid_search_180_w3_1}
  \par\medskip 
  \includegraphics[width=\textwidth]{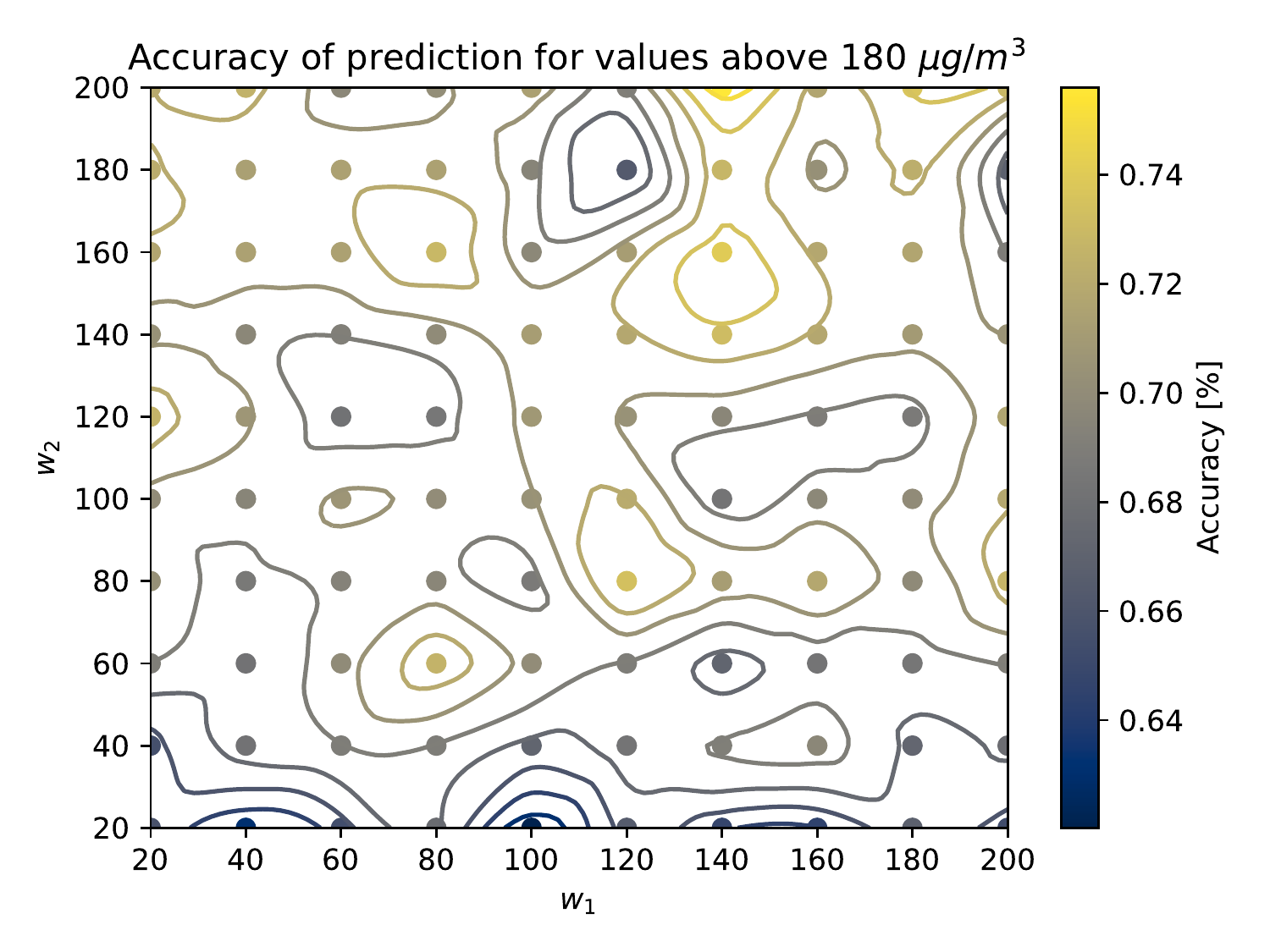}
  \caption{Optimization of classes [5, 6]\\$w_3=20$}  \label{fig:grid_search_180_w3_2}
\end{subfigure}
\hspace*{\fill}
\begin{subfigure}{0.32\textwidth}
  \includegraphics[width=\textwidth]{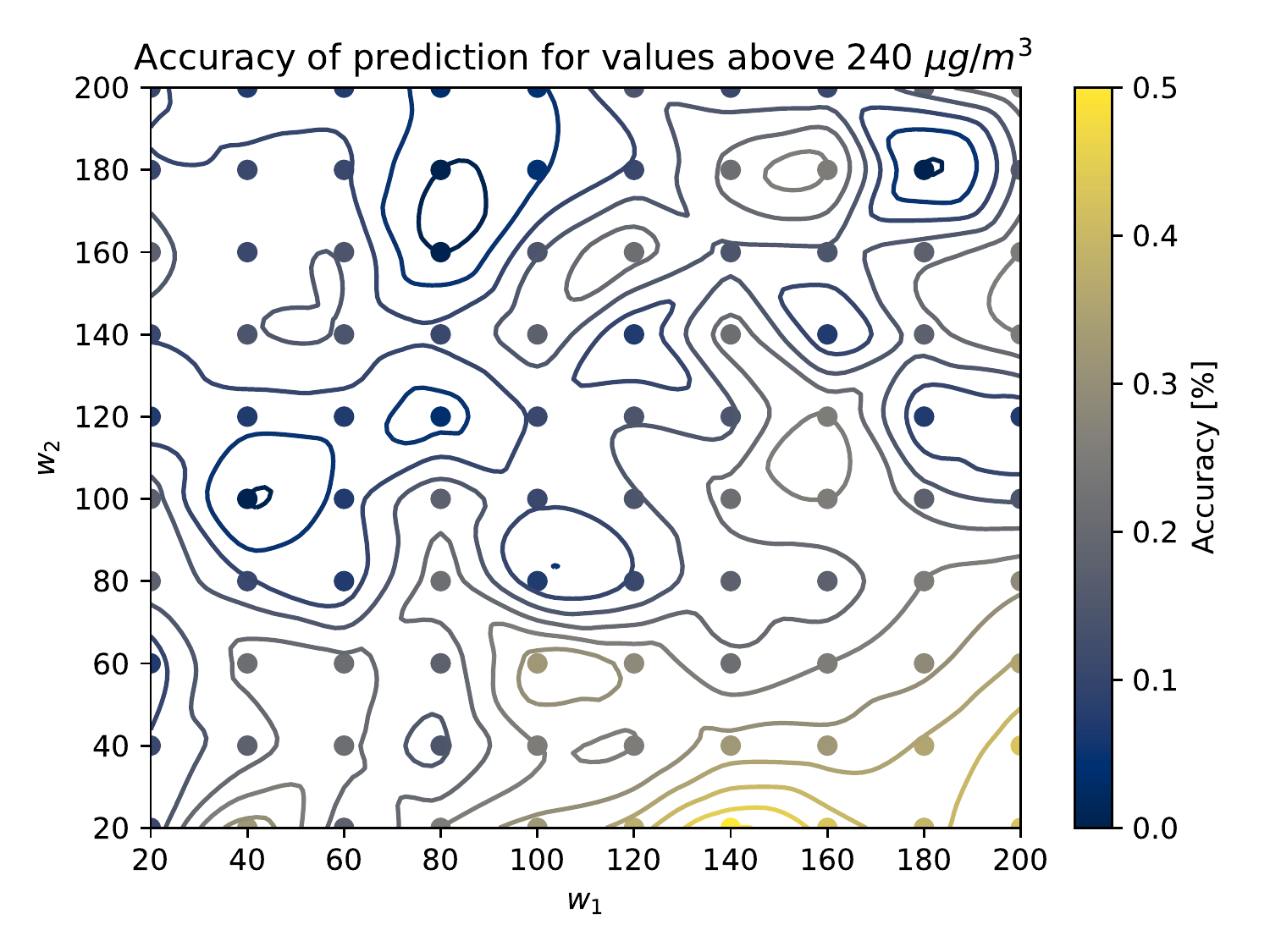}
  \caption{Optimization of class 6\\$w_3=10$} 
  \label{fig:grid_search_240_w3_1}
  \par\medskip 
  \includegraphics[width=\textwidth]{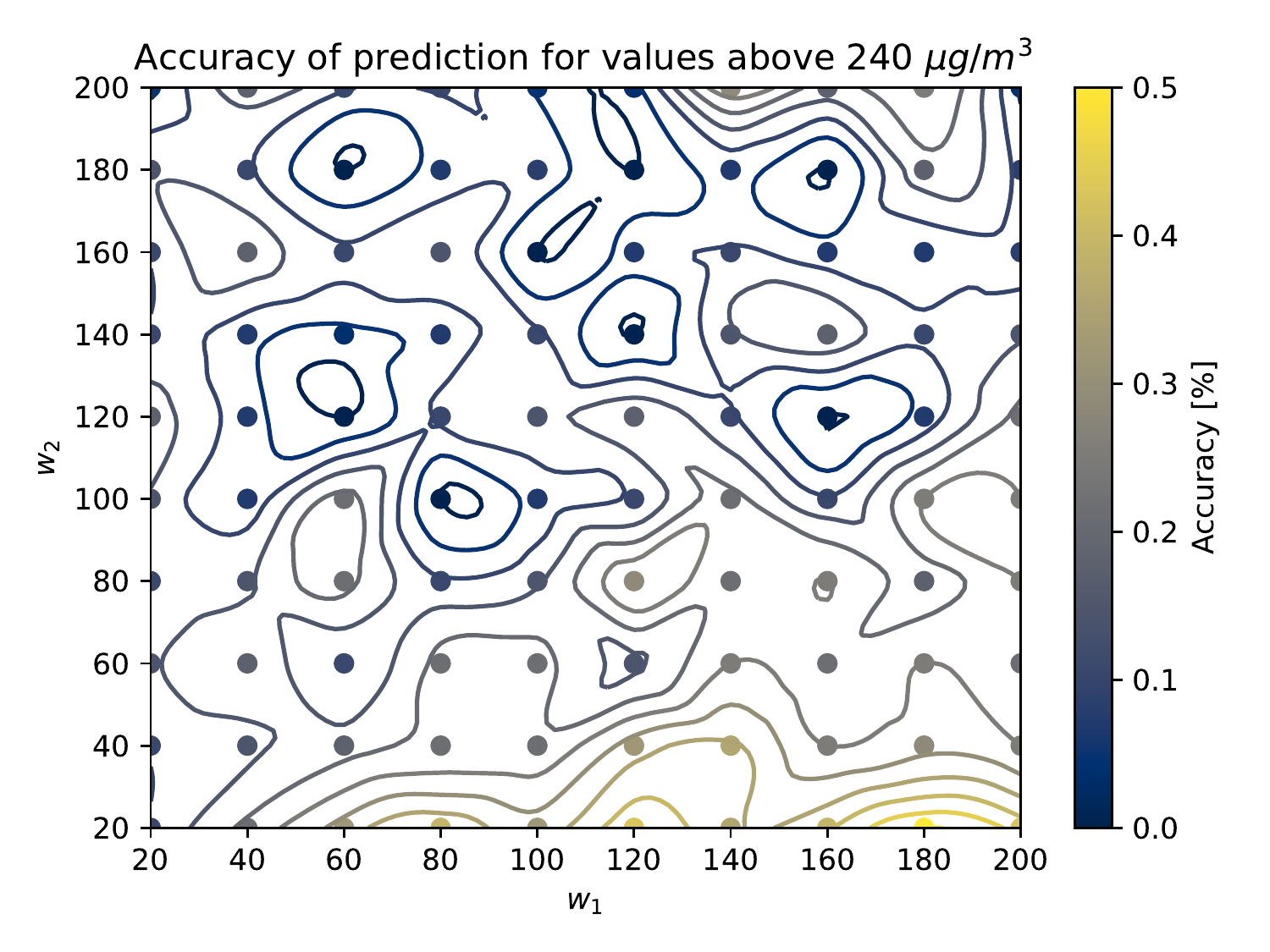}
  \caption{Optimization of class 6\\$w_3=20$}  \label{fig:grid_search_240_w3_2}
\end{subfigure}
\caption{Plot of the results of the weights grid search with 100 selected features, aggregated across the stations of Chiasso, Bioggio, Mendrisio and Locarno}
\label{fig:grid_search}
\end{figure*}

\begin{figure*}
    \centering
        \begin{subfigure}[b]{0.24\textwidth}
            \includegraphics[width=\textwidth]{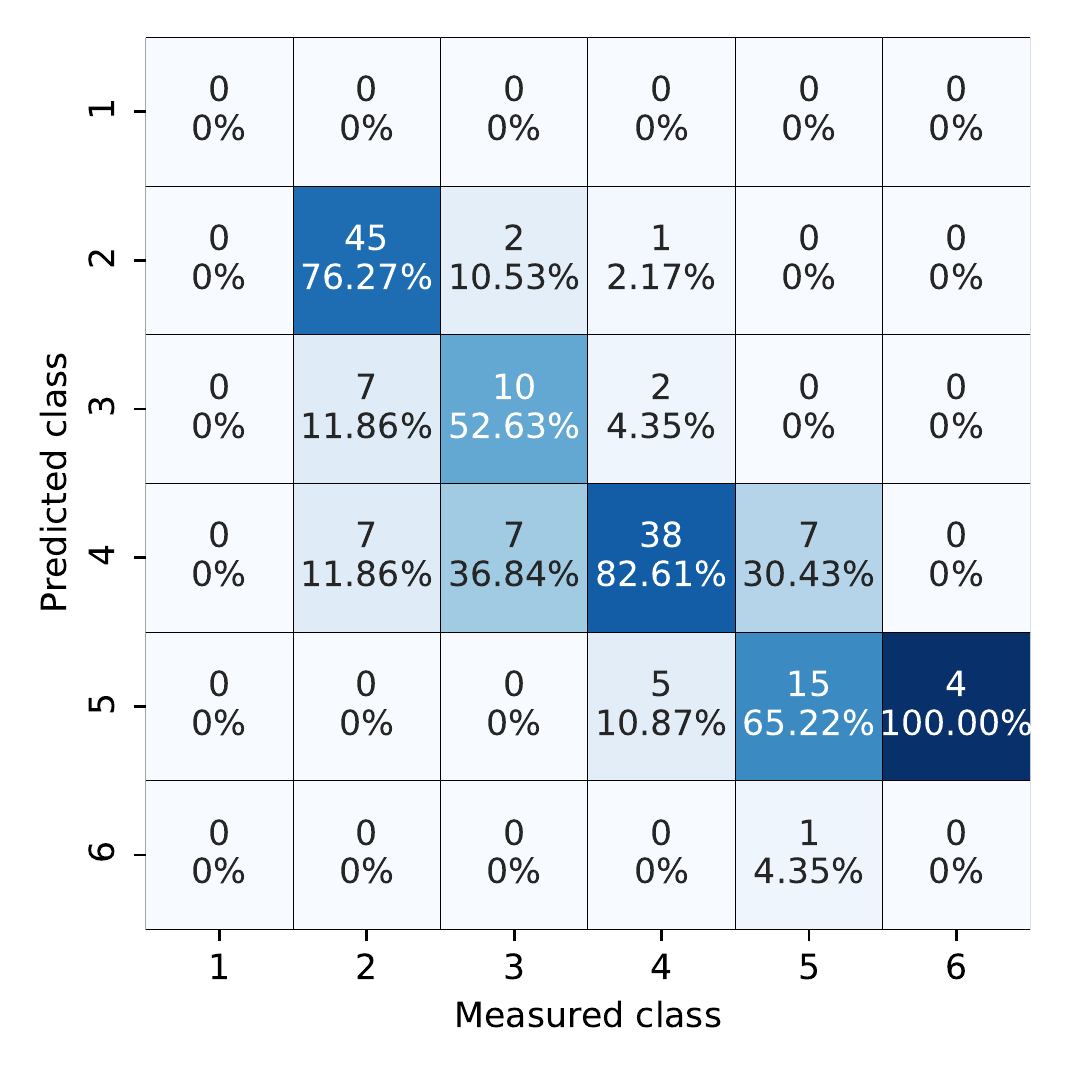}
            \caption{No weights applied}
        \end{subfigure}
        \begin{subfigure}[b]{0.24\textwidth}
            \includegraphics[width=\textwidth]{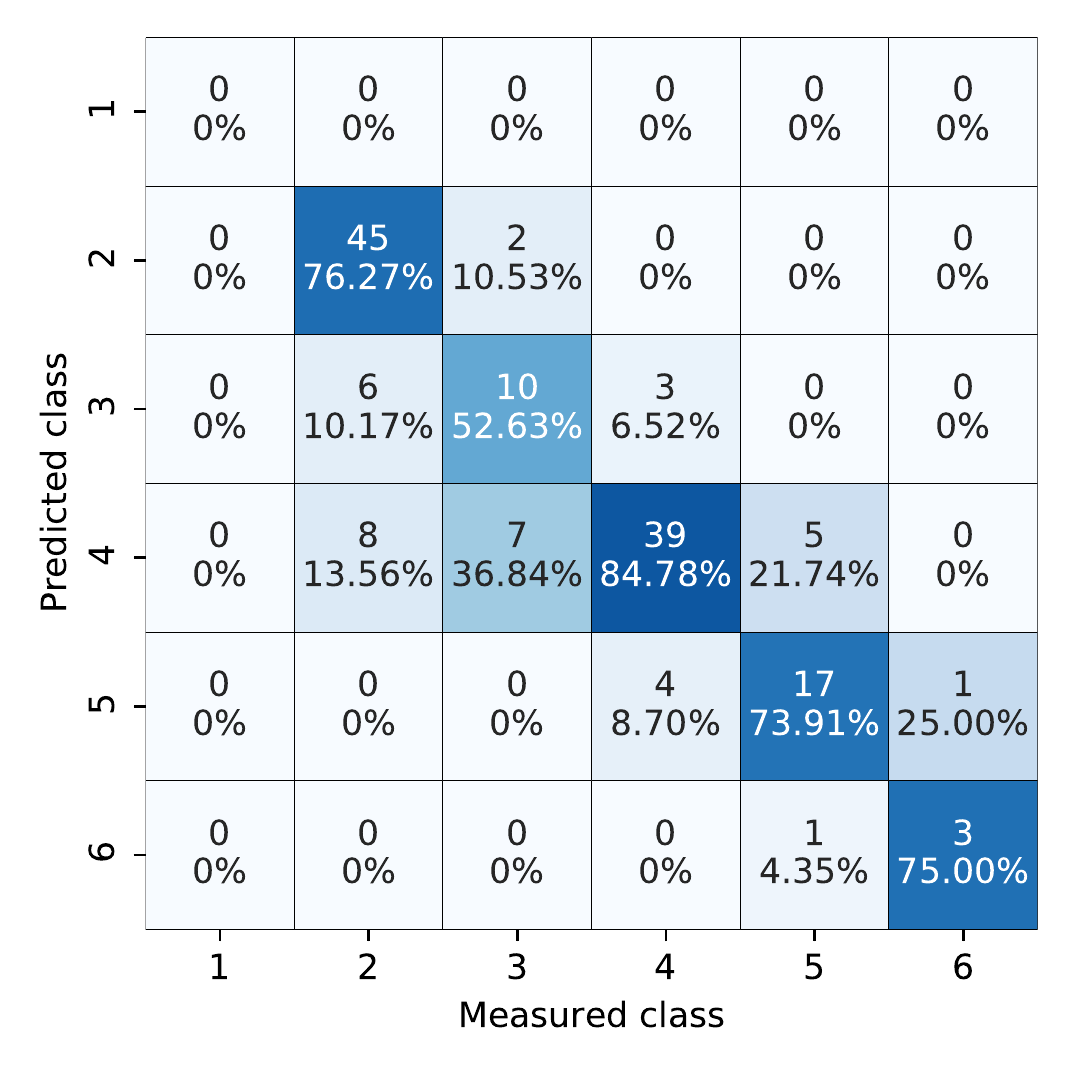}
            \caption{Optimizing classes [4, 5, 6]}
        \end{subfigure}
        \begin{subfigure}[b]{0.24\textwidth}
            \includegraphics[width=\textwidth]{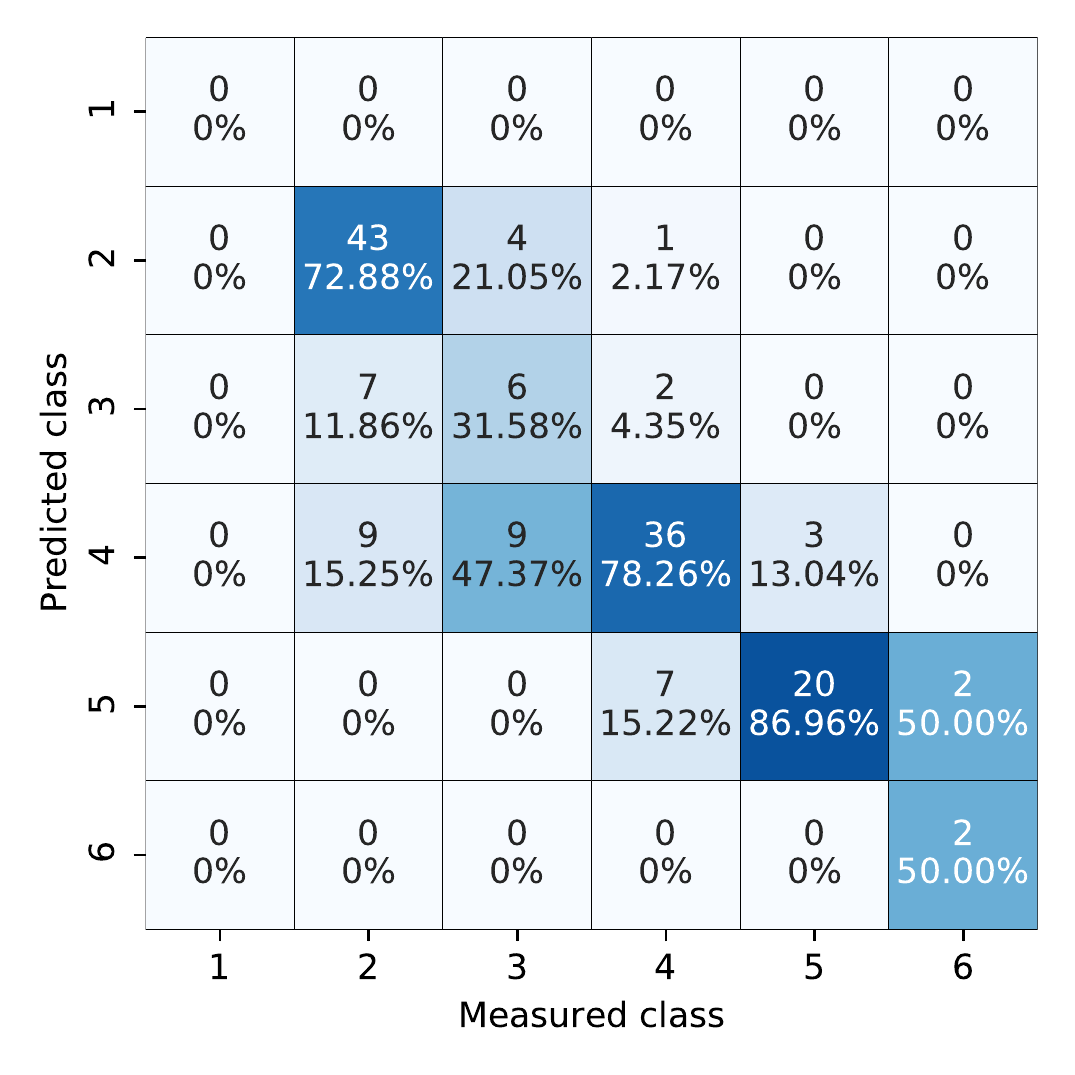}
            \caption{Optimizing classes [5, 6]}
        \end{subfigure}
        \begin{subfigure}[b]{0.24\textwidth}
            \includegraphics[width=\textwidth]{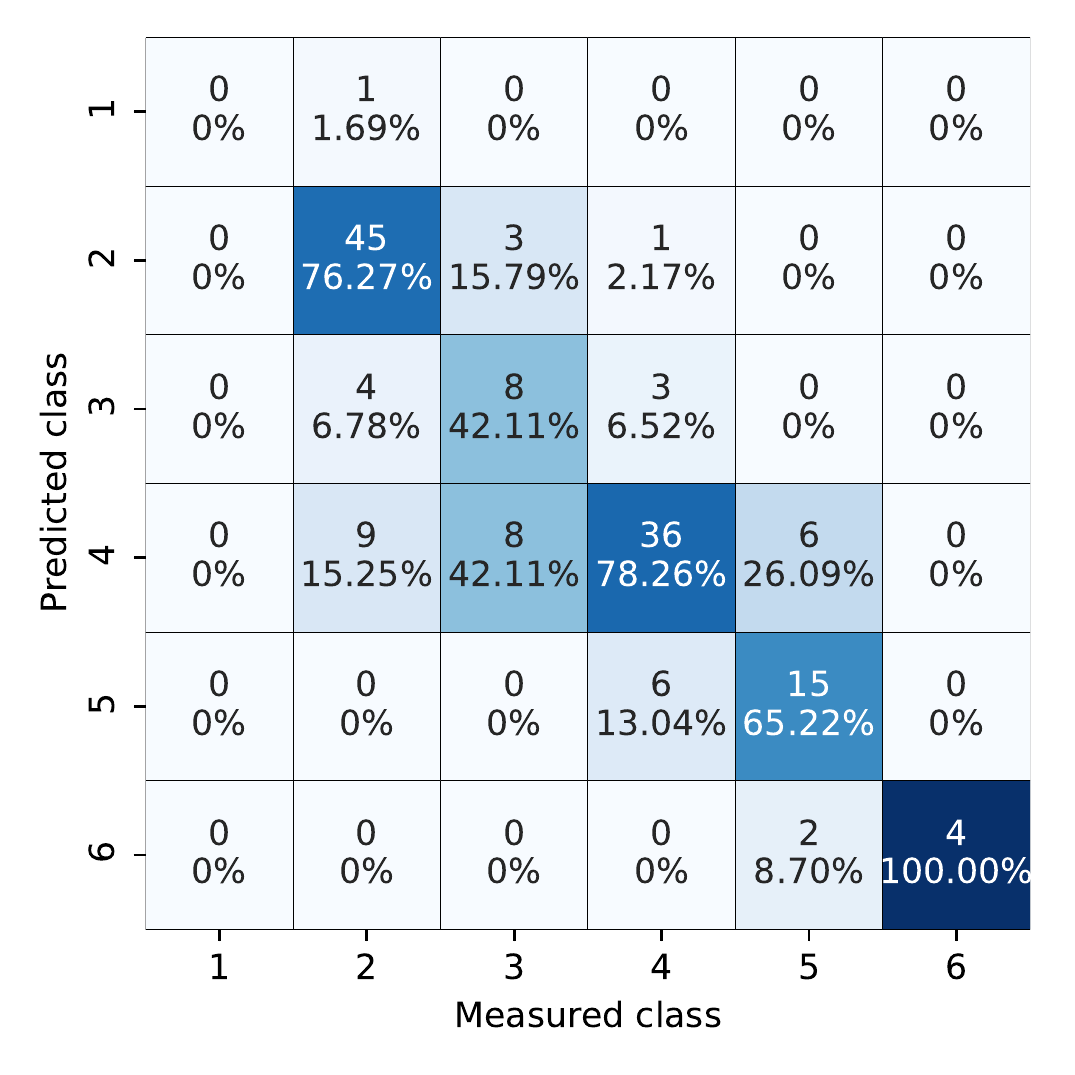}
            \caption{Optimizing class 6}
        \end{subfigure}
        \caption{Confidence matrices for Chiasso \textit{MOR}}    
        \label{fig:confMats}
\end{figure*}

We analyzed the four stations with the highest number of extreme measurements in 2019: Bioggio, Mendrisio, Locarno, and Chiasso. All these stations registered at least one event of class 6 and many events of class 5, as shown in Table \ref{tab:events}. In particular, Chiasso registered 27 measurements above 180 $\mu g/m^3$, of which 4 above 240 $\mu g/m^3$. 
In contrast to what we observed with unweighted training, we noticed that when using weighted training, increasing the number of selected features above 30 improves the prediction accuracy at high ozone concentrations. Therefore, we decided to increase the number of features that the algorithm can use to perform its prediction to 100. We used SHAP as feature selection method and NGBoost as regressor. We calculated the KPIs of the models for each combination of the weights $w$ with $w_1,w_2\in\{2,4,6,\ldots,20\}$, $w_3\in\{1,2\}$.  Fig. \ref{fig:grid_search} shows the aggregated accuracy of the prediction for the three different classes of interest. Fig. \ref{fig:grid_search_135_w3_1} and  \ref{fig:grid_search_135_w3_2} illustrate the distribution of the accuracy when considering only the ozone measured values above 135 $\mu g/m^3$. Similarly, Fig. \ref{fig:grid_search_180_w3_1}, \ref{fig:grid_search_180_w3_2} and \ref{fig:grid_search_240_w3_1}, \ref{fig:grid_search_240_w3_2} show the accuracy when restricting ourselves only to values above 180 and 240 $\mu g/m^3$ respectively. The continuous iso-lines are obtained with cubic interpolation. It is difficult to infer which weights give the best results, especially when trying to maximise the accuracy of observations above 135 $\mu g/m^3$, but for the other 2 cases, high $w_2$ and variable $w_1$ give the best weights for observations above 180 $\mu g/m^3$, while high $w_1$ and low $w_2$ appear to be the best weights combination for correctly predicting observations above 240 $\mu g/m^3$.\\
Table \ref{tab:results_weight} shows the results of weighted training for the considered stations. We report the KPIs and the three sets of weights that gave the best accuracy for each fraction of the dataset, compared to the results obtained by the case where no weights are applied. We can see that the introduction of the weights does not unduly affect the KPIs, which in fact improve in some cases. For Chiasso \textit{MOR}, we also show in Figure \ref{fig:confMats} the complete confusion matrices obtained. It can be seen that when actively trying to enhance recognition of observations above 135 $\mu g/m^3$, the correct classification of these values boosted from about 50-75\% to about 80-85\%. Similarly, for the observations above 180 $\mu g/m^3$ the correct recognition rises from 50-70\% to 80-90\% in all stations but Locarno, where it stops at 70\%. Finally, in Mendrisio and Chiasso, we could correctly predict all values above 240 $\mu g/m^3$, which was not achieved in the unweighted analysis. This is not the case for Bioggio and Locarno, where the only class 6 value is never recognized.\\

\begin{table*}
\centering
\setlength{\tabcolsep}{0.7em} 
{\renewcommand{\arraystretch}{1.2}
\caption{Results of the weighted analysis}
\label{tab:results_weight}
\begin{tabular}{|c|l|c|c|c|c|c|c|c|c|c|}
\hline
\multicolumn{2}{|l|}{\multirow{2}{*}{}}                                                                     & \multicolumn{3}{c|}{\textbf{Weights}} & \multicolumn{3}{c|}{\textbf{Target Class Accuracy}}                & \multicolumn{3}{c|}{\textbf{KPIs}} \\ \cline{3-11} 
\multicolumn{2}{|l|}{}                                                                                      & $w_1$          & $w_2$         & $w_3$         & cl. $\left[4,5,6\right]$ & cl. $\left[5,6\right]$ & cl. 6 & MAE      & RMSE     & Tot. Acc.   \\ \hline
\multirow{4}{*}{\vtop{\hbox{\strut BIO}\hbox{\strut MOR}}} & No weights                                                              & -           & -          & -          & 72.72                & 61.11                & 0.00                 & 13.28    & 17.66    & 70.95        \\ \cline{2-2}
                                  & \begin{tabular}[c]{@{}l@{}}max. accuracy classes $\left[4,5,6\right]$\end{tabular} & 14          & 8          & 2          & 85.07                & 77.78                & 0.00                 & 13.03    & 17.52    & 73.68        \\ \cline{2-2}
                                  & \begin{tabular}[c]{@{}l@{}}max. accuracy classes $\left[5,6\right]$\end{tabular} & 12          & 20         & 1          & 79.10                & 88.89                & 0.00                 & 13.61    & 18.25    & 71.05        \\ \cline{2-2}
                                  & \begin{tabular}[c]{@{}l@{}}max. accuracy class 6\end{tabular} & 18          & 10         & 1          & 77.61                & 72.22                & 0.00                 & 12.62    & 16.18    & 70.39        \\ \hline
\multirow{4}{*}{\vtop{\hbox{\strut BIO}\hbox{\strut EVE}}} & No weights                                                              & -           & -          & -          & 69.69                & 61.11                & 0.00                 & 14.22    & 18.50    & 68.71        \\ \cline{2-2}
                                  & \begin{tabular}[c]{@{}l@{}}max. accuracy classes $\left[4,5,6\right]$\end{tabular} & 4           & 10         & 2          & 84.84                & 77.78                & 0.00                 & 14.59    & 18.47    & 73.51        \\ \cline{2-2}
                                  & \begin{tabular}[c]{@{}l@{}}max. accuracy classes $\left[5,6\right]$\end{tabular} & 6           & 18         & 1          & 83.33                & 83.33                & 0.00                 & 15.08    & 19.27    & 69.53        \\ \cline{2-2}
                                  & \begin{tabular}[c]{@{}l@{}}max. accuracy class 6\end{tabular} & 8           & 4          & 1          & 83.33                & 72.22                & 0.00                 & 14.24    & 18.31    & 73.50        \\ \hline
\multirow{4}{*}{\vtop{\hbox{\strut CHI}\hbox{\strut MOR}}} & No weights                                                              & -           & -          & -          & 72.60                & 55.56                & 0.00                 & 14.61    & 18.79    & 71.52        \\ \cline{2-2}
                                  & \begin{tabular}[c]{@{}l@{}}max. accuracy classes $\left[4,5,6\right]$\end{tabular} & 18          & 2          & 1          & 80.82                & 74.07                & 75.00                & 13.75    & 19.54    & 75.50        \\ \cline{2-2}
                                  & \begin{tabular}[c]{@{}l@{}}max. accuracy classes $\left[5,6\right]$\end{tabular} & 14          & 20         & 2          & 79.45                & 81.48                & 50.00                & 15.51    & 21.33    & 70.86        \\ \cline{2-2}
                                  & \begin{tabular}[c]{@{}l@{}}max. accuracy class 6\end{tabular} & 20          & 4          & 1          & 75.34                & 70.37                & 100.00               & 13.53    & 19.14    & 71.52        \\ \hline
\multirow{4}{*}{\vtop{\hbox{\strut CHI}\hbox{\strut EVE}}} & No weights                                                              & -           & -          & -          & 65.75                & 48.15                & 0.00                 & 15.28    & 20.43    & 67.35        \\ \cline{2-2}
                                  & \begin{tabular}[c]{@{}l@{}}max. accuracy classes $\left[4,5,6\right]$\end{tabular} & 8           & 2          & 2          & 82.19                & 74.07                & 75.00                & 14.90    & 19.96    & 71.52        \\ \cline{2-2}
                                  & \begin{tabular}[c]{@{}l@{}}max. accuracy classes $\left[5,6\right]$\end{tabular} & 20          & 20         & 1          & 75.34                & 85.19                & 75.00                & 15.26    & 20.13    & 73.51        \\ \cline{2-2}
                                  & \begin{tabular}[c]{@{}l@{}}max. accuracy class 6\end{tabular} & 18          & 2          & 2          & 75.34                & 70.37                & 75.00                & 15.26    & 20.09    & 67.55        \\ \hline
\multirow{4}{*}{\vtop{\hbox{\strut MEN}\hbox{\strut MOR}}} & No weights                                                              & -           & -          & -          & 77.14                & 68.18                & 50.00                & 14.69    & 19.49    & 71.62        \\ \cline{2-2}
                                  & \begin{tabular}[c]{@{}l@{}}max. accuracy classes $\left[4,5,6\right]$\end{tabular} & 12          & 4          & 2          & 84.29                & 81.82                & 100.00               & 14.17    & 19.02    & 71.43        \\ \cline{2-2}
                                  & \begin{tabular}[c]{@{}l@{}}max. accuracy classes $\left[5,6\right]$\end{tabular} & 18          & 6          & 1          & 80.00                & 86.36                & 100.00               & 14.56    & 19.40    & 67.35        \\ \cline{2-2}
                                  & \begin{tabular}[c]{@{}l@{}}max. accuracy class 6\end{tabular} & 20          & 4          & 1          & 77.14                & 81.82                & 100.00               & 14.34    & 19.23    & 69.39        \\ \hline
\multirow{4}{*}{\vtop{\hbox{\strut MEN}\hbox{\strut EVE}}} & No weights                                                              & -           & -          & -          & 74.28                & 72.72                & 50.00                & 15.29    & 19.38    & 68.02        \\ \cline{2-2}
                                  & \begin{tabular}[c]{@{}l@{}}max. accuracy classes $\left[4,5,6\right]$\end{tabular} & 12          & 12         & 2          & 82.86                & 81.82                & 50.00                & 16.13    & 21.13    & 67.81        \\ \cline{2-2}
                                  & \begin{tabular}[c]{@{}l@{}}max. accuracy classes $\left[5,6\right]$\end{tabular} & 20          & 6          & 1          & 74.29                & 90.91                & 100.00               & 16.53    & 21.06    & 64.38        \\ \cline{2-2}
                                  & \begin{tabular}[c]{@{}l@{}}max. accuracy class 6\end{tabular} & 18          & 4          & 1          & 81.43                & 81.82                & 100.00               & 15.54    & 19.75    & 67.81        \\ \hline
\multirow{4}{*}{\vtop{\hbox{\strut LOC}\hbox{\strut MOR}}} & No weights                                                              & -           & -          & -          & 54.90                & 40.00                & 0.00                 & 14.12    & 18.45    & 64.86        \\ \cline{2-2}
                                  & \begin{tabular}[c]{@{}l@{}}max. accuracy classes $\left[4,5,6\right]$\end{tabular} & 18          & 10         & 2          & 80.39                & 60.00                & 0.00                 & 14.43    & 18.52    & 72.11        \\ \cline{2-2}
                                  & \begin{tabular}[c]{@{}l@{}}max. accuracy classes $\left[5,6\right]$\end{tabular} & 4           & 14         & 1          & 72.55                & 70.00                & 0.00                 & 15.06    & 19.65    & 68.03        \\ \cline{2-2}
                                  & \begin{tabular}[c]{@{}l@{}}max. accuracy class 6\end{tabular} & 6           & 18         & 1          & 70.59                & 70.00                & 0.00                 & 13.92    & 17.97    & 70.07        \\ \hline
\multirow{4}{*}{\vtop{\hbox{\strut LOC}\hbox{\strut EVE}}} & No weights                                                              & -           & -          & -          & 52.94                & 40.00                & 0.00                 & 14.09    & 18.84    & 65.99        \\ \cline{2-2}
                                  & \begin{tabular}[c]{@{}l@{}}max. accuracy classes $\left[4,5,6\right]$\end{tabular} & 4           & 8          & 2          & 80.39                & 50.00                & 0.00                 & 14.37    & 18.61    & 69.18        \\ \cline{2-2}
                                  & \begin{tabular}[c]{@{}l@{}}max. accuracy classes $\left[5,6\right]$\end{tabular} & 6           & 20         & 1          & 76.47                & 70.00                & 0.00                 & 14.71    & 19.83    & 67.12        \\ \cline{2-2}
                                  & \begin{tabular}[c]{@{}l@{}}max. accuracy class 6\end{tabular} & 10          & 4          & 1          & 66.67                & 50.00                & 0.00                 & 13.93    & 17.75    & 67.81        \\ \hline
\end{tabular}}
\end{table*}

\section{Conclusions} \label{sec:conclusions}
In this paper, we forecasted the day-ahead maximum of the ground-level ozone concentration during the summer period of 2019 in 7 localities located in Southern Switzerland, using a physics-agnostic, data-driven approach. Due to the high number of signals potentially affecting the predictions, we performed a preliminary feature selection using two methods, which we compared. The selected features were then used to train different state-of-the-art forecasting algorithms. Analyzing feature importance interactions using Shapley values suggests that the models trained through our learning pipeline effectively learned explanatory cross-dependencies among atmospheric variables, which are described in the ozone photochemistry literature. \\
Our analysis showed that Gradient Boosting algorithms, and in particular Least Square Boosting and Natural Gradient Boosting, consistently performed better than the other tested forecasting methods. Where possible, we further compared our results with those of other papers, and we were able to conclude our results are similar to previous analysis and, in some cases, even better.\\
We then evaluated the effect of weighted training to increase the accuracy of predictions for high ozone concentrations. Our analysis shows that this method is feasible, as it increases forecast accuracy without compromising overall forecast quality. Future directions for this work include the formulation of probabilistic techniques for the robust estimation of annual ozone concentration peaks, which are the most difficult events to predict, due to their scarcity in the training set. In this view, training the forecasters with ad-hoc generated adversarial examples could result in a better forecast of the conditional probability distributions.
\FloatBarrier
\onecolumn{
\twocolumn
\bibliographystyle{elsarticle-harv}
\bibliography{references}}

\end{document}